\documentclass{article}
\usepackage[preprint]{neurips_2021}
\usepackage[utf8]{inputenc} 
\usepackage[T1]{fontenc}    
\usepackage{hyperref}       
\usepackage{url}            
\usepackage{booktabs}       
\usepackage{amsfonts}       
\usepackage{nicefrac}       
\usepackage{microtype}      
\usepackage{xcolor}         
\usepackage{multicol}
\usepackage{multirow}
\usepackage{wrapfig}

\usepackage{amsmath,amsthm,mathtools,amsfonts,amssymb}
\usepackage{graphicx}
\usepackage{algorithm, algorithmic}
\usepackage{subfigure,subfloat}
\usepackage{soul}
\usepackage{csquotes}

\title{Controllable seismic velocity synthesis using generative diffusion models}

\author{%
Fu Wang$^{1}$,~Xinquan Huang$^{1}$,~Tariq Alkhalifah$^{1}$\\
$^1$King Abdullah University of Science and Technology\\
\texttt{\{fu.wang,xinquan.huang,tariq.alkhalifah\}@kaust.edu.sa}\\
}

\begin{document}
\maketitle

\title{Controllable seismic velocity synthesis using generative diffusion models}

\begin{abstract}
Accurate seismic velocity estimations are vital to understanding Earth's subsurface structures, assessing natural resources, and evaluating seismic hazards. Machine learning-based inversion algorithms have shown promising performance in regional (i.e., for exploration) and global velocity estimation, while their effectiveness hinges on access to large and diverse training datasets whose distributions generally cover the target solutions. Additionally, enhancing the precision and reliability of velocity estimation also requires incorporating prior information, e.g., geological classes, well logs, and subsurface structures, but current statistical or neural network-based methods are not flexible enough to handle such multi-modal information. To address both challenges, we propose to use conditional generative diffusion models for seismic velocity synthesis, in which we readily incorporate those priors. This approach enables the generation of seismic velocities that closely match the expected target distribution, offering datasets informed by both expert knowledge and measured data to support training for data-driven geophysical methods. We demonstrate the flexibility and effectiveness of our method through training diffusion models on the OpenFWI dataset under various conditions, including class labels, well logs, reflectivity images, and the combination of these priors. The performance of the approach under out-of-distribution conditions further underscores its generalization ability, showcasing its potential to provide tailored priors for velocity inverse problems and create specific training datasets for machine learning-based geophysical applications.
\end{abstract}

\section{Introduction}
Full waveform inversion (FWI) plays a vital role in creating detailed subsurface models by repeatedly matching observed seismograms on the Earth surface or within a well with wave-equation-based simulated data \cite[]{Tarantola1984}. 
However, FWI faces significant challenges due to the constraints that lead to incomplete observations, such as limited bandwidth or limited sensors. These limitations often result in less than optimal, and sometimes even unsuccessful, seismic velocity estimation, all while incurring high computational costs. 
The advent of modern deep learning techniques opens new possibilities for addressing these challenges in seismic velocity estimation, offering potential improvements in both accuracy and efficiency.
In recent years, deep learning-based methods have received considerable attention from the geophysical community as it has been used in seismic data processing \cite[]{yu2019deep,wang2019deep,wang2019intelligent,saad2020deep,harsuko2022storseismic}, inversion\cite[]{yang2019deep,li2019deep,wu2019inversionnet,chen2020automatic,sun2021physics,sun2023implicit,wang2023prior}, and interpretation \cite[]{zhao2018seismic,wu2019faultseg3d,di2020seismic,wang2022seismic}. However, compared to the other two tasks, applying deep learning to inverse problems, specifically direct velocity inversion, is particularly challenging though efficient. 
It requires that neural networks (NNs) are well-trained to learn the complex non-linear mapping between the observed data and its corresponding subsurface models. 
To achieve accurate instant inversion, we need offline training of the neural network model on massive training data to ensure the accuracy and generalization of the NNs. In addition to the quantity, the distribution of the training data needs to be close to that corresponding to real applications, to ensure good results \cite[]{kazei2020velocity}. However, acquiring reasonable velocity models for training such direct inversion is not trivial, and requires a lot of human intervention. There are some open datasets for velocities and seismic data, e.g., SEAM dataset \cite[]{fehler2008seg}, US east coast deep water line 32 \cite[]{klitgord1994geophysical}, USGS Marine seismic data \cite[]{triezenberg2016national}, SEG/EAGE Salt and Overthrust models \cite[]{aminzadeh19963}, BP dataset \cite[]{billette20052004}. However, those datasets were originally prepared to evaluate numerical full waveform inversion and interpretation as the amount of data is limited for a potential training of a deep neural network-based inversion. 

Specifically designed datasets like the OpenFWI and OpenFWI2 datasets, as well as deterministic ways to build diverse velocity models \cite[]{ren2021building,liu2021deep} for machine learning-based velocity inversion, have recently emerged. However, to scale the neural networks to accommodate real-world scenarios, the datasets often used possess several limitations: 1) They rely on manually defined velocities, which lack the necessary diversity for large-scale neural networks. 2) Synthetic datasets demand significant physical space for storage and sharing. In comparison to datasets explicitly tailored for the task, deep learning can be employed to generate velocities. \cite{ovcharenko2019style} proposed to use a style transfer approach to generate velocities based on synthetic prior velocities. Alternatively, generative models, which have shown considerable potential in capturing the distribution of a dataset and enabling the generation of new samples, can be a more efficient data synthesis tool. Generative models have favorable advantages: 1) they can produce high-quality and diverse large datasets as their learnable parameters are used to store the distribution; 2) they provide potentially unlimited training data from the stored distribution features; 3) they can utilize both OpenFWI datasets as well as other datasets (SEAM, Overthrust), even velocity models from industry, to generate more realistic velocities; 4) using the samples from the generative model for training can further improve the performance of NNs \cite[]{he2022synthetic}. 
\cite{puzyrev2022geophysical} proposed to use GANs to generate velocity models, but training GANs is fraught with difficulties, and their progress in generative tasks remains uncertain.
In addition, the learned velocity distributions— leveraging these generative models— often lack essential foundational information regarding the velocity model classes (general geological structural description like salt, fault, flat, curve) or features, rendering them somewhat blind. 
In this paper, we use conditional diffusion models to synthesize seismic velocity models. 
It can provide desired and rich velocities, dependent on the target velocity classes, well logs, and structural information, as training datasets for data-driven geophysical methods. 

Beyond addressing the challenge of lacking training datasets whose distribution covers the target solution, another important challenge we often face is the inflexibility of incorporating multi-modal priors into FWI.
Along these lines, \cite{wang2023prior} demonstrated that injecting priors stored in diffusion models into the inversion process can improve the results. This is done by regularly updating the velocity model using diffusion denoising regularizer after each iteration, leading to increased accuracy of the seismic velocity model.
However, the priors used in their approach mainly provide high-level information, reflecting general velocity ranges and structures. While useful, this is often insufficient for achieving the desired level of accuracy in scenarios of poorly acquired data. 
In this case, additional multi-modal prior information (geological classes, well logs, and structure information) is needed. 
Hence, the critical challenge lies in how to handle these multi-modal priors flexibly, and that can be addressed by controllable velocity synthesis. 

Thus, in this paper, we introduce a conditional diffusion model to control the generation process of the velocity model, where we can use the conditional sampling of diffusion models with fruitful conditions such as well log information, velocity classes, or detailed subsurface structures (migration profiles or reflectivity images). 
Specifically, we use the classifier-free guidance and a cross-attention mechanism for the conditional diffusion model. Compared to the reconstruction-guided sampling method, which is an approach based on a classifier-guided method, the proposed method can better control the velocity synthesis. We summarize the contributions of this paper in the following:

\begin{itemize}
    \item We propose to use conditional diffusion models for velocity synthesis.    
    \item We demonstrate the framework for controllable velocity synthesis with three types of conditions, velocity classes, well-log information, and reflectivity images.  
    \item Numerical examples on OpenFWI datasets show that we can control the velocity using any type of condition or all of them to generate high-quality velocities. 
    \item We show that the generation is even credible for out-of-distribution conditions.
\end{itemize}

The rest of the paper is organized as follows: we first review the concept of denoising diffusion probabilistic models in Section~\ref{method} and then introduce ways to incorporate the prior conditions, e.g., the class labels, smoothness of the velocity, well condition, as well as imaging results, into the generation process. To showcase the feasibility of the proposed method in velocity generation, we present numerical experiments in Section~\ref{exp}. Finally, we discuss the features and limitations of the approach and state our conclusions in Section~\ref{conclusion}

\section{Methology}
\label{method} 
Generative models are widely used in seismic exploration, such as in denoising and interpolating seismic data \cite[]{oliveira2018interpolating,oliveira2019improving,gatti2020towards, LI2020seismic, Feng2021, wei2021reconstruction, wei2023seismic}, seismic inversion and imaging \cite[]{mosser2020stochastic, meng2021seismic, normal2021,saraiva2021data, wang2023prior,yin2023wise,baldassari2024conditional}, interpretation \cite[]{10.1190/geo2019-0438.1,durall2021generative,10.1190/tle40070534.1}, and so on. These generative models mainly belong to four types of generative frameworks: variational autoencoders (VAEs) \cite[]{kingma2013auto}, generative adversarial networks (GANs) \cite[]{goodfellow2014generative}, normalizing flows \cite[]{dinh2014nice}, and diffusion models \cite[]{sohl-dickstein2015deep}. Among all generative models, VAE and normalizing flows require stringent constraints on the model architecture or resort to surrogate objectives for approximating the maximum likelihood training. As mentioned earlier, training GANs is fraught with difficulties. Recently, diffusion models, which circumvent the above limitations, have beaten GANs in generative quality and diversity \cite[]{dhariwal2021diffusion}. Here, we choose diffusion models for seismic velocity synthesis. Specifically, to achieve the controllable generation, we use conditional diffusion models. 

\begin{figure}[h]
    \centering
    \includegraphics[width=0.9\textwidth]{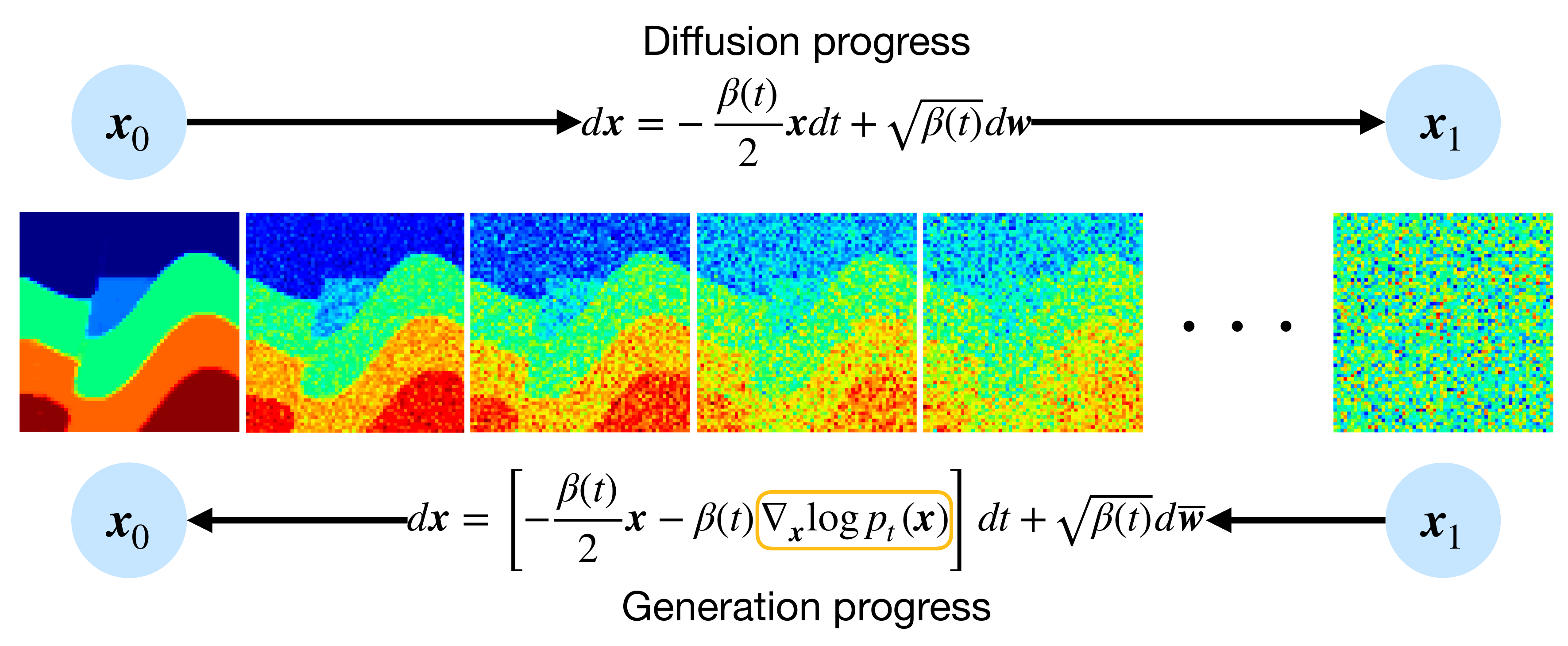}
    \caption{An overview of the generative diffusion model. \textbf{Top}: forward process; \textbf{Bottom}: unconditional sampling with a score function $\nabla_{\boldsymbol{x}} \log p_t\left(\boldsymbol{x}\right)$.}
    \label{fig:digram1}
\end{figure}

The fundamental concept behind Denoising Diffusion Probabilistic Models (DDPMs) is straightforward: we aim to acquire the ability to restore images in the 2D context (or videos in the 3D realm) by training denoisers to reverse the corruption process, ultimately restoring the original data distribution from a Gaussian distribution of the latent space. This comprehensive framework can be delineated into two distinct stages: the training phase and the sampling phase, as depicted in Figure~\ref{fig:digram1}, where we use seismic velocities as single-channel images. In real-world applications, we aim to exert control over the velocity generation through conditional factors. This control can be introduced during the training process, which is referred to as `classifier-free guidance', or it can be implemented directly in the sampling stage, known as the `classifier-guided (reconstruction-guided) diffusion model'. In this section, we will initially delve into the training stage before exploring the sampling process in subsequent subsections.
\subsection{Training process}
During the training, given clean data (seismic velocity models) $\boldsymbol{x} \sim p(\boldsymbol{x})$, we use a forward process (Gaussian process) to corrupt the data and obtain the noisy $\boldsymbol{x}_t$, $\boldsymbol{x}_t \sim p_t(\boldsymbol{x})$, which satisfies a linear stochastic differential equation (SDE) \citep{song2020score}:
\begin{equation}
d \boldsymbol{x}=-\frac{\beta(t)}{2} \boldsymbol{x} d t+\sqrt{\beta(t)} d \boldsymbol{w}, t \in[0,1], 
\label{equ:sde}
\end{equation}
where $\beta(t)$ is a differentiable noise scheduler to enforce $\boldsymbol{x}_1$ to satisfy a pre-defined noise distribution (here it is the standard Gaussian distribution), $\boldsymbol{w}$ is the Wiener process (Brownian motion, which adds random noise to data at each time step), and $d$ denotes the change in a variable, like $\boldsymbol{x}$, $t$, or $\boldsymbol{w}$ in Equation~\ref{equ:sde}. As shown at the top of Figure~\ref{fig:digram1}, for a clean data sample $\boldsymbol{x}_0$, we gradually add noise to the data $\boldsymbol{x}_t$ until we obtain a final image $\boldsymbol{x}_1$ of random noise. 
The discrete form of Equation~\ref{equ:sde} (taking DDPM as an example) can be formulated as:
\begin{equation}
    \boldsymbol{x}_i=\sqrt{\bar{\alpha}_i} \boldsymbol{x}_0+\sqrt{1-\bar{\alpha}_i} \boldsymbol{\epsilon}, \boldsymbol{\epsilon} \sim \mathcal{N}(0, \boldsymbol{I}), i=1,\dots, N
\end{equation}
where $\bar{\alpha}_i=\prod_{j=1}^i (1-\beta_j)$. Then, we use a neural network $\Phi(\theta)$ to learn to denoise the $\boldsymbol{x}_i$ into an estimate $\Phi(\theta, \boldsymbol{x}_i)\approx \boldsymbol{x}$ for all discrete time points $N$.  The loss function can be defined by means of a weighted mean square error loss
\begin{equation}
\mathbb{E}_{\boldsymbol{\epsilon}, t}\left[ 
\left(\frac{1}{1 - \bar{\alpha}_i}\right)^{(-\gamma)}
\left\|\Phi\left(\theta,\boldsymbol{x}_i\right)-\boldsymbol{x}\right\|_2^2\right],
\end{equation}
where $\gamma$ is the loss weight value \cite[]{choi2022perception}. In practice, we do not use the direct prediction of $\boldsymbol{x}$ via the neural networks. 
Instead, we use the $\boldsymbol{\epsilon}$-prediction parameterization, which often results in better sample quality \citep{ho2020denoising}, defined as $\Phi(\theta,\boldsymbol{x}_i)=(\frac{\boldsymbol{x}_i}{\sqrt{\bar{\alpha}_i}} - \sqrt{\frac{1-\bar{\alpha}_i}{\bar{\alpha}_i}}\boldsymbol{\epsilon}_\theta(\boldsymbol{x}_i))$ and train the $\boldsymbol{\epsilon}_\theta$ in $\boldsymbol{\epsilon}$ space.
It results in the solution of a scaled score estimate $\boldsymbol{\epsilon}_\theta(z_t)\approx -\sqrt{1-\bar{\alpha}_t}\nabla_{\boldsymbol{x}} \log p_t\left(\boldsymbol{x}\right)$.

\subsection{Unconditional sampling process}
Unconditional sampling in diffusion models involves generating new samples from the model's learned distribution without any specific conditioning. This is different from conditional sampling, which we will discuss in the next subsection. Conditional sampling entails the model being conditioned on specific input data to generate samples that align with that input.

To generate an unconditional sample, a diffusion model starts with a random Gaussian noisy image and then applies a sequence of denoising steps to reduce the noise gradually, as shown at the bottom of Figure~\ref{fig:digram1}. 
At each step, the model predicts a residual image that is added to the noisy image to produce a less noisy image. 
The process continues until the noise is sufficiently reduced and a high-quality sample is obtained. The sampling process can be described in the continuous form as solving a reverse-time SDE \citep{song2020score}
\begin{equation}
d \boldsymbol{x}=\left[-\frac{\beta(t)}{2} \boldsymbol{x}-\beta(t) \nabla_{\boldsymbol{x}} \log p_t(\boldsymbol{x})\right] d t+\sqrt{\beta(t)} d \overline{\boldsymbol{w}},
    \label{equ:sampling_con}
\end{equation}
where $\nabla$ represents the gradient operation, $\overline{\boldsymbol{w}}$ is the reverse-time standard Wiener process, and $\nabla_{\boldsymbol{x}} \log p_t(\boldsymbol{x})$
is the score function of the distribution $p_t(\boldsymbol{x})$ learned by a neural network. In the implementation, we take the discrete form of Equation~\ref{equ:sampling_con} as follows:
\begin{equation}
\boldsymbol{x}_{i-1}=\frac{1}{\sqrt{\alpha_i}}\left(\boldsymbol{x}_i-\frac{1-\alpha_i}{\sqrt{1-\bar{\alpha}_i}} \boldsymbol{\epsilon}_\theta\left(\boldsymbol{x}_i\right)\right)+\sqrt{\frac{\beta_i(\alpha_i-\bar{\alpha}_i)}{\alpha_i(1-\bar{\alpha}_i)}}\boldsymbol{\epsilon}, \quad \boldsymbol{\epsilon} \sim \mathcal{N}(0, \boldsymbol{I}).
    \label{equ:sampling_dis}
\end{equation}

One of the advantages of unconditional sampling in diffusion models is that it can be used to generate samples that are diverse and creative. This is because the latent space of the diffusion model is as large as the image, and the path to diverse images includes a lot of (like 1000) steps. Also, in the unconditional form, the model is not constrained to generate samples that match any specific input criterion. Instead, the model is free to explore the full range of possible outputs that are consistent with the learned distribution. However, in many realistic scenarios, we hope to control the velocity generation based on given conditions, e.g., the class label of the subsurface, the well information, the smoothness, or the imaging results. In the next subsection,  we will introduce how to incorporate the condition into the generation process.

\subsection{Conditional sampling process}
In the conditional sampling (conditional generation) setting, the generation process can be formulated in the context that given the conditioning signal $\boldsymbol{c}$, which can represent a class label, smoothness level, well information, imaging results, or any other type of condition. The model generates output $\boldsymbol{x}$ satisfying the distribution of $p(\boldsymbol{x})$ while consistent with the set condition. Two conditional samplers are introduced here, including the reconstruction-guided samplers and classifier-free guidance samplers. The former can be applied to any pre-trained diffusion model without additional training cost, as the guidance only happens at the sampling stage, while the latter requires training the diffusion model to include the conditions.

\textbf{Reconstruction guided generation}. In this way, our target is to make the generated velocity from the unconditional generation sampler $\Phi(\theta,\boldsymbol{x})$ satisfy the conditions, in which we can use an objective function to guide the process at each time step $i$:
\begin{equation}
    L = \left\|\boldsymbol{c}-mask(\Phi\left(\theta, \boldsymbol{x}_i\right))\right\|_2^2,
\end{equation}
where $mask$ denotes the mask operation on the generated samples to extract, for our specific example, the well location information to compare with the condition given by the true well log. 
We are using this objective function to calculate the gradient with respect to $\boldsymbol{x_i}$ to update the sampler $\Phi(\theta, x_i)$. 
Then, the final samples at each time step $t$ is 
\begin{equation}
    \hat{\Phi}(\theta, \boldsymbol{x_i}) \leftarrow \Phi(\theta, \boldsymbol{x}_i) - \frac{w_r \sqrt{\bar{\alpha}_i}}{2} \nabla_{\boldsymbol{x}_i}\left\|\boldsymbol{c}-mask(\Phi\left(\theta, \boldsymbol{x}_i\right))\right\|_2^2,
\end{equation}
where the additional term with reweighting in this equation compared to the original sampler $\Phi(\theta,\boldsymbol{x}_i)$ is the guidance based on the model's reconstruction of the condition, so-called reconstruction-guided sampling \cite[]{ho2022video}, $
w_r$ is the weighting factor to control the strength of the condition, the left arrow to denote the replacement of the original unconditional sampling procedure $\Phi(\theta, \boldsymbol{x}_i)$ by the conditional sampling implementation $\hat{\Phi(\theta, \boldsymbol{x}_i)}$.  
With this implementation, the trajectory of the reverse process is modified by the gradient of the objective function to guide the generated velocity to satisfy the prior distribution as well as the given well condition.
However, this process is very sensitive to the choice of the weighting factor, which we will demonstrate via the numerical results, yielding inflexible control. 

\textbf{Classifier-free guidance sampler.} 
When considering the classes of the images, a concept similar to the reconstruction-guided diffusion model is called classifier guidance diffusion, whose generation process is guided by the gradient from an external classifier. Although it can feasibly guide the unconditional diffusion model with a plug-in classifier, this process will increase the computational cost as it requires gradient calculation during the inference. Another drawback of this type of method is that the generation sometimes will ignore the guidance of the classifier via the adversarial noise, and then the generation, in this case, will actually be unconditional. Another controllable sampling is classifier-free guidance sampling. Compared to reconstruction-guided sampling, this requires training the diffusion model with the condition as part of the training. The good news is that it often has higher-quality generation during the sampling stage. Our training target is to train a neural network to fit the distribution $p(\boldsymbol{x}|\boldsymbol{c})$ given a conditioning signal $\boldsymbol{c}$. Compared to the previously mentioned diffusion model (Figure~\ref{fig:digram1}), the only modification is changing the $\Phi(\theta,\boldsymbol{x})$ to $\Phi(\theta,\boldsymbol{x},\boldsymbol{c})$. To improve the sample quality, instead of directly using the prediction $\Phi(\theta,\boldsymbol{x},\boldsymbol{c})$, we can use the following weighting form with the ability to adjust the strength of the conditioning,
\begin{equation}
    \hat{\boldsymbol{\epsilon}}_\theta(\boldsymbol{x}_i,\boldsymbol{c}) = (1+\lambda)\boldsymbol{\epsilon}_\theta(\boldsymbol{x}_i,\boldsymbol{c}) - \lambda \boldsymbol{\epsilon}_\theta(\boldsymbol{x}_i),
\end{equation}
where $\lambda$ is the conditioning scale, and $\Phi(\theta,\boldsymbol{x}_i,\boldsymbol{c})=(\frac{\boldsymbol{x}_i}{\sqrt{\bar{\alpha}_i}} - \sqrt{\frac{1-\bar{\alpha}_i}{\bar{\alpha}_i}}\boldsymbol{\epsilon}_\theta(\boldsymbol{x}_i,\boldsymbol{c}))$ is the conditional model prediction. When $\lambda$ is positive, the model output will have lower diversity, but higher quality (high-fidelity) compared to the case where the output is only dependent on the $\boldsymbol{\epsilon}_\theta(\boldsymbol{x}_i,\boldsymbol{c})$ \citep{ho2022classifier}. To make the diffusion model more flexible, in training, we can randomly add embedding vectors of the condition as well as the vectors of a null condition denoting unconditioning, yielding flexible generation abilities when sampling. With this trick in the training stage, the unconditioned $\boldsymbol{\epsilon}_\theta(\boldsymbol{x}_i)$ can be realized by setting $\boldsymbol{c}$ as a vector of null condition in $\boldsymbol{\epsilon}_\theta(\boldsymbol{x}_i,\boldsymbol{c})$.
Next, we will discuss how to incorporate the condition in the neural networks.
\begin{figure}
    \centering
    \includegraphics[width=0.8\textwidth]{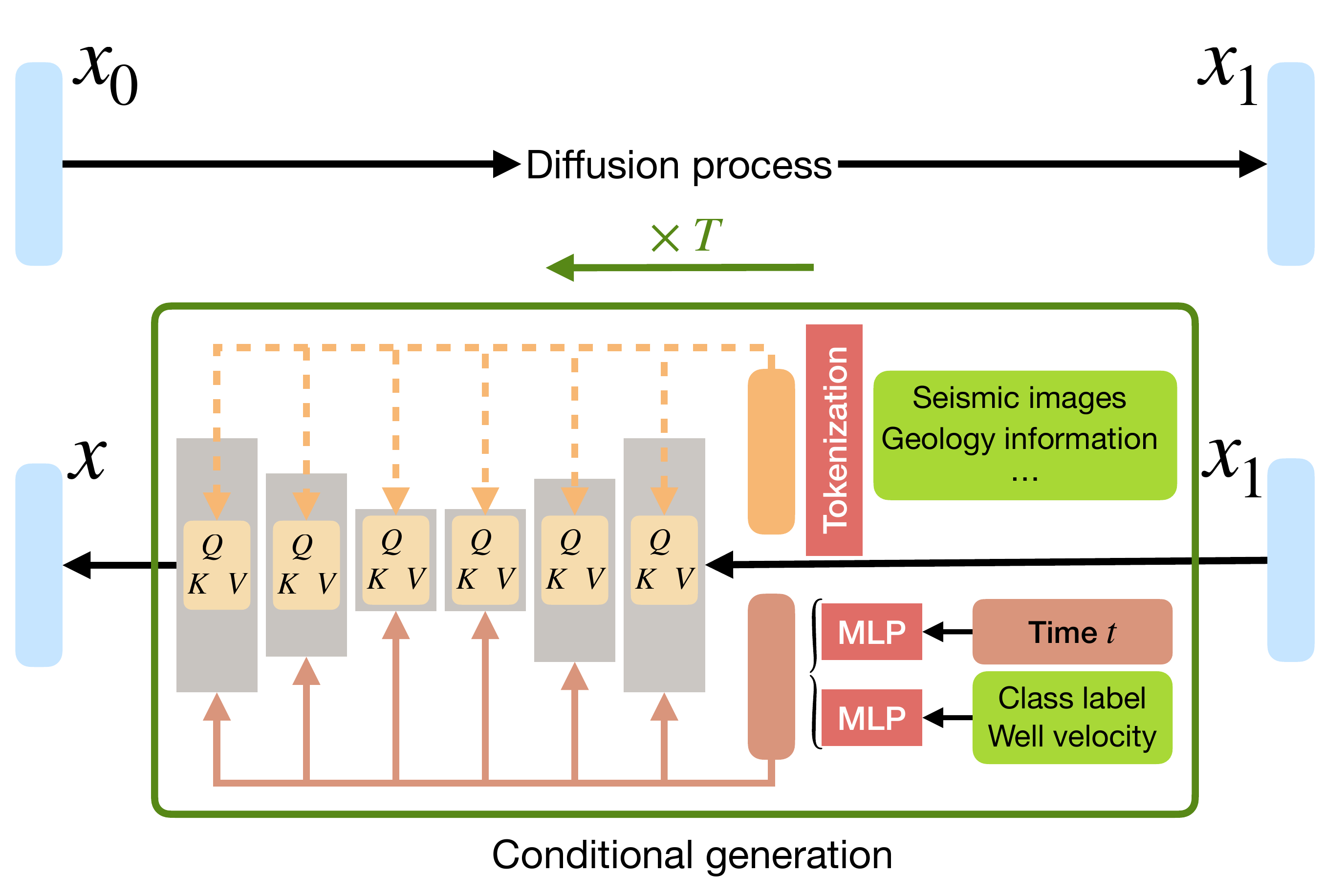}
    \caption{Conditional generative diffusion models. The backbone network is U-net \citep{ronneberger2015u}.}
    \label{fig:digram2}
\end{figure}

As shown in Figure~\ref{fig:digram2}, there are two ways to achieve the classifier-free guidance. In the first way, the class label or the well log information is embedded as a vector and then added to the hidden layer of the U-Net. Specifically, for the class labels, we start an embedding layer to convert the class to a vector, followed by a Multilayer perceptron (MLP) to obtain an embedded vector while for the well log, we use an MLP to embed the well logs and sinusoidal positional encoding to embed the well location, yielding the embedded vectors.
However, these MLPs have limits in handling high-dimensional conditions, e.g., multiple wells or seismic images. In the second way, for 2D input data for conditioning, we utilize cross-attention to add conditional information into the neural networks for training. The concept is based on calculating the similarity between the conditioning data and the hidden output of the U-Net and then updating the latent vectors in the U-Net.
We map the tokenizations (specifically for seismic velocity, it is a way to transform the image to the embedded vectors) of the original conditions $\mathbf{c}$, $\tau_{\theta}(\mathbf{c})$, to the intermediate output of the U-Net by implementing cross-attention$(Q, K, V)=softmax(\frac{QK^T}{\sqrt{d}})\cdot V$, between the U-Net features and the input condition \citep{rombach2022high}, where
\begin{equation}
    Q=W^Q_{j} \cdot \phi_j\left(\boldsymbol{x}_i\right), K=W^K_{j} \cdot \tau_{\theta}(\boldsymbol{c}), V=W^V_{j} \cdot \tau_{\theta}(\boldsymbol{c}).
    \label{equ:attention}
\end{equation}
Here, $\phi_j(\boldsymbol{x}_i)$ is the output of the intermediate layer of the U-Net, $\tau_{\theta}$ used in this paper is $1\times1$ convolutional layer, and $W_j^Q$, $W_j^K$, and $W_j^V$ are the learnable weight matrices for the query, key, and value, respectively. 
The cross-attention mechanism enables the model to consider and incorporate information from given conditions while processing the velocity synthesis. 
To achieve an accurate control of velocity synthesis, we add the spatial information of the velocity (the coordinates of the grid points) as additional channels and then concatenate them with conditions $\boldsymbol{c}$ to feed into the $\tau_{\theta}$.
The beauty of this method is that it can incorporate multi-model conditioning. What's more, thanks to the cross-attention, we can achieve pixel-level controllable generation. 


\section{Numerical Experiments}
\label{exp}
\begin{figure}[h]
    \centering
    \includegraphics[width=0.75\textwidth]{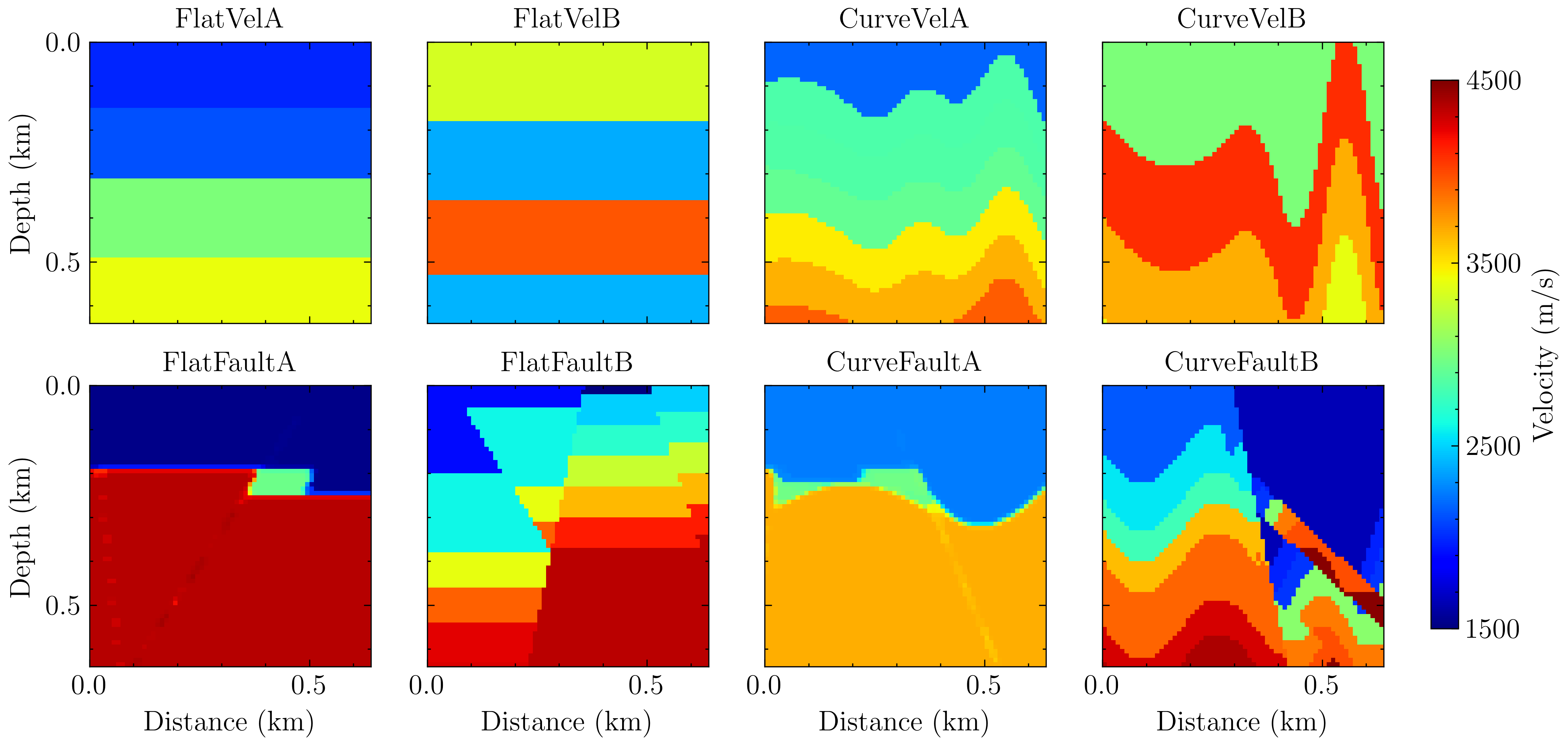}
    \caption{Samples of velocity models corresponding to the various classes.}
    \label{fig:vel_example}
\end{figure}

\begin{figure}[h]
    \centering
    \includegraphics[width=0.75\textwidth]{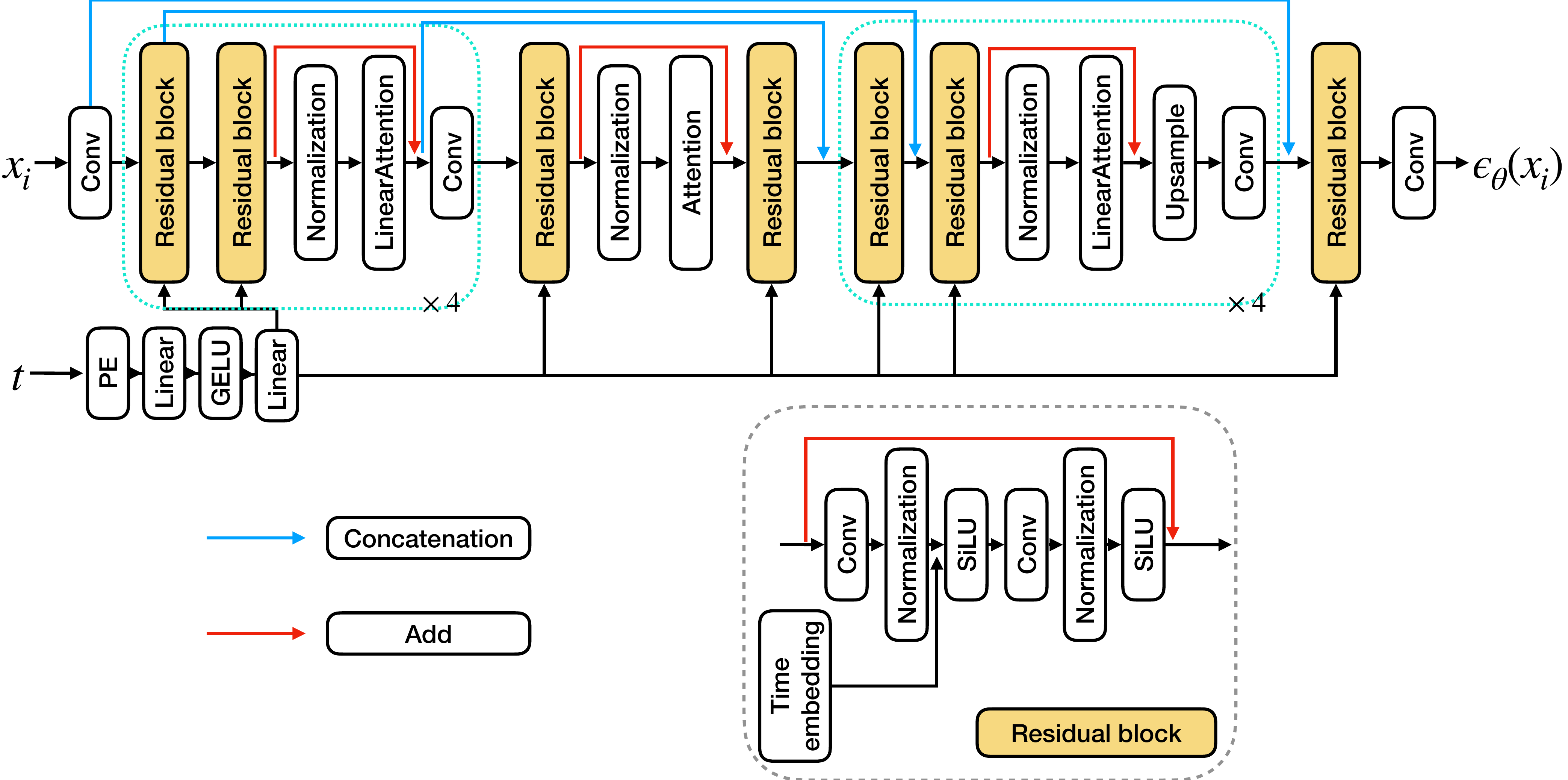}
    \caption{The overall architecture of the $
    \boldsymbol{\epsilon}$-prediction network. Conv denotes a convolutional layer and PE denotes a sinusoidal positional encoding. The dashed box below shares the architecture of a residual block.}
    \label{fig:network}
\end{figure}

The proposed method can handle a wide range of multi-modal conditions to control the velocity synthesis. 
Here, we report our tests on the OpenFWI \cite[]{deng2021openfwi} dataset and showcase the flexibility of the proposed method to incorporate different conditions, specifically the class labels, well logs, and reflectivity images, to control the velocity synthesis. We utilize the 2D velocity models in these tests, which include eight classes of models, so-called \enquote{FlatVel-A}, \enquote{FlatVel-B}, \enquote{CurveVel-A}, \enquote{CurveVel-B}, \enquote{FlatFault-A}, \enquote{FlatFault-B}, \enquote{CurveFault-A}, \enquote{CurveFault-B}, with samples shown in Figure~\ref{fig:vel_example}. 
They have distinct features in structures and velocity variations with depth. \enquote{FlatVelA} and \enquote{CurveVelA} represent models without fault features, displaying a gradual increase in velocity with depth. In contrast, \enquote{FlatVelB} and \enquote{CurveVelB} also lack fault features, but here the velocity distribution is random along the depth profile. 
\enquote{FlatFaultA} and \enquote{CurveFaultA} introduce fault features with fewer discontinuities and more subtle variations in velocity. These are in contrast to \enquote{FlatFaultB} and \enquote{CurveFaultB}, where fault features are present with more discontinuities and more pronounced velocity changes.
There are 336000 velocity samples in the dataset, and the model size of each velocity in our test is 64$\times$64 and the spatial interval is 10 m in horizontal and vertical directions. We train the diffusion model using an Adam optimizer with a learning rate of 1e-4. The batch size for the training is 1024, and the maximum training iteration is set to 200000. The backbone of the diffusion model is a U-Net with attention blocks \cite[]{oktay2018attention}. We use four feature map resolutions (from 64$\times$64 to 8$\times$8). The diffusion timesteps, $t$, is set to 1000 and embedded via a sinusoidal positional encoding into each residual block. The details of the backbone network architecture are shown in Figure~\ref{fig:network}. When applying cross-attention to incorporate the conditions, we replace the normalization and attention block in Figure~\ref{fig:network} into a spatial transformer.  
\subsection{Unconditional Denoising Diffusion Probability models}
\begin{figure}[h]
    \centering
    \includegraphics[width=1.0\textwidth]{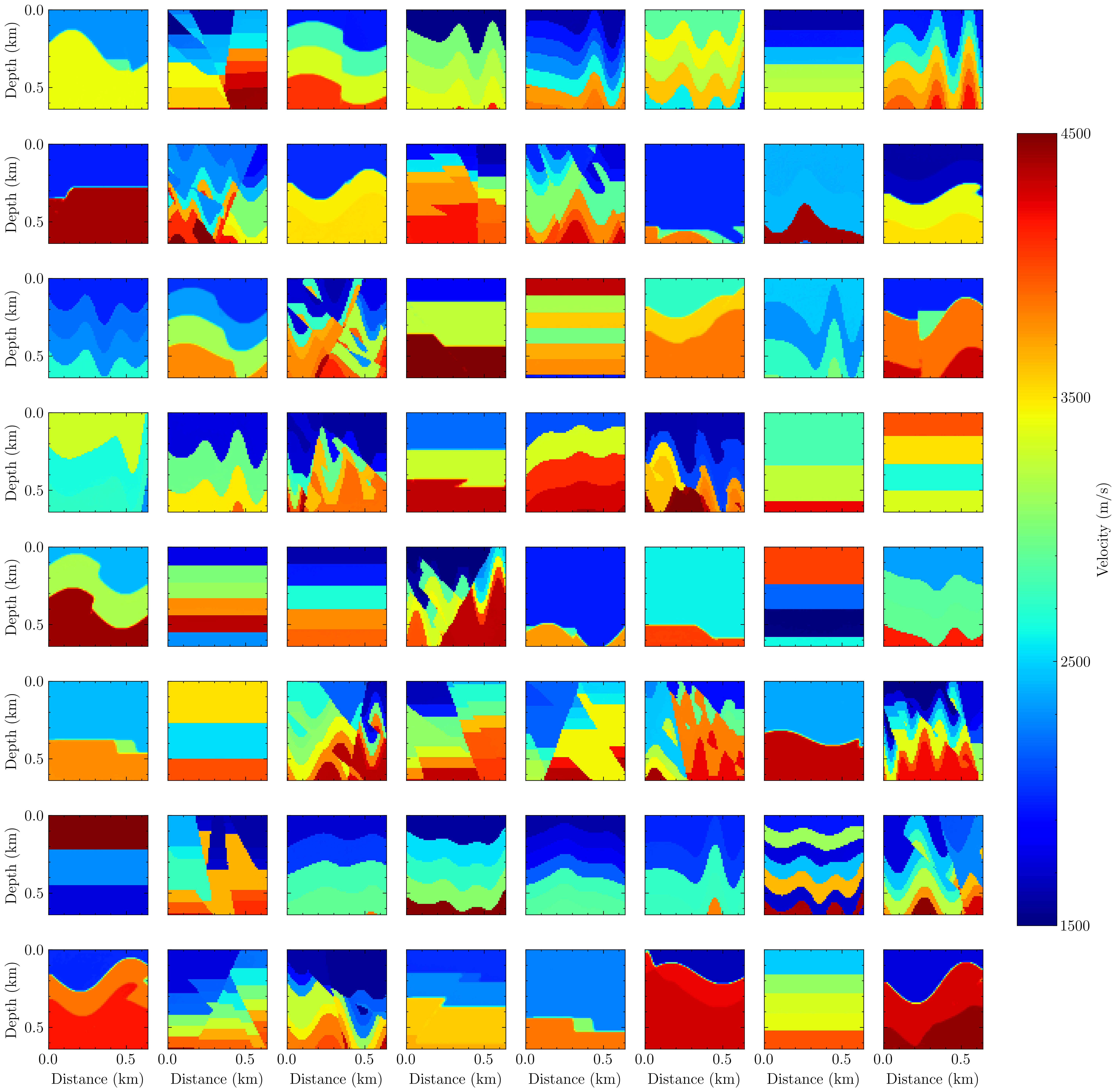}
    \caption{The generated samples by diffusion models without conditioning.}
    \label{fig:unconditional_generation}
\end{figure}

In such a setting, the training process is focused on the denoising neural networks. We are using the U-Net model to learn the distribution $p(x)$. We evaluate the generation progress with 64 random Gaussian noise samples as input to the reverse progress, and Figure~\ref{fig:unconditional_generation} shows the corresponding generated velocities. Without imposing any condition to affect the sampling process, the diffusion model produces velocities with the maximum diversity, which means the generated velocities can belong to any of the eight previously mentioned classes, or even be in a form that represents a combination of the classes, depending on their distribution.

\subsection{Class-conditioned generation}
\begin{figure}[h]
    \centering
    \includegraphics[width=1.0\textwidth]{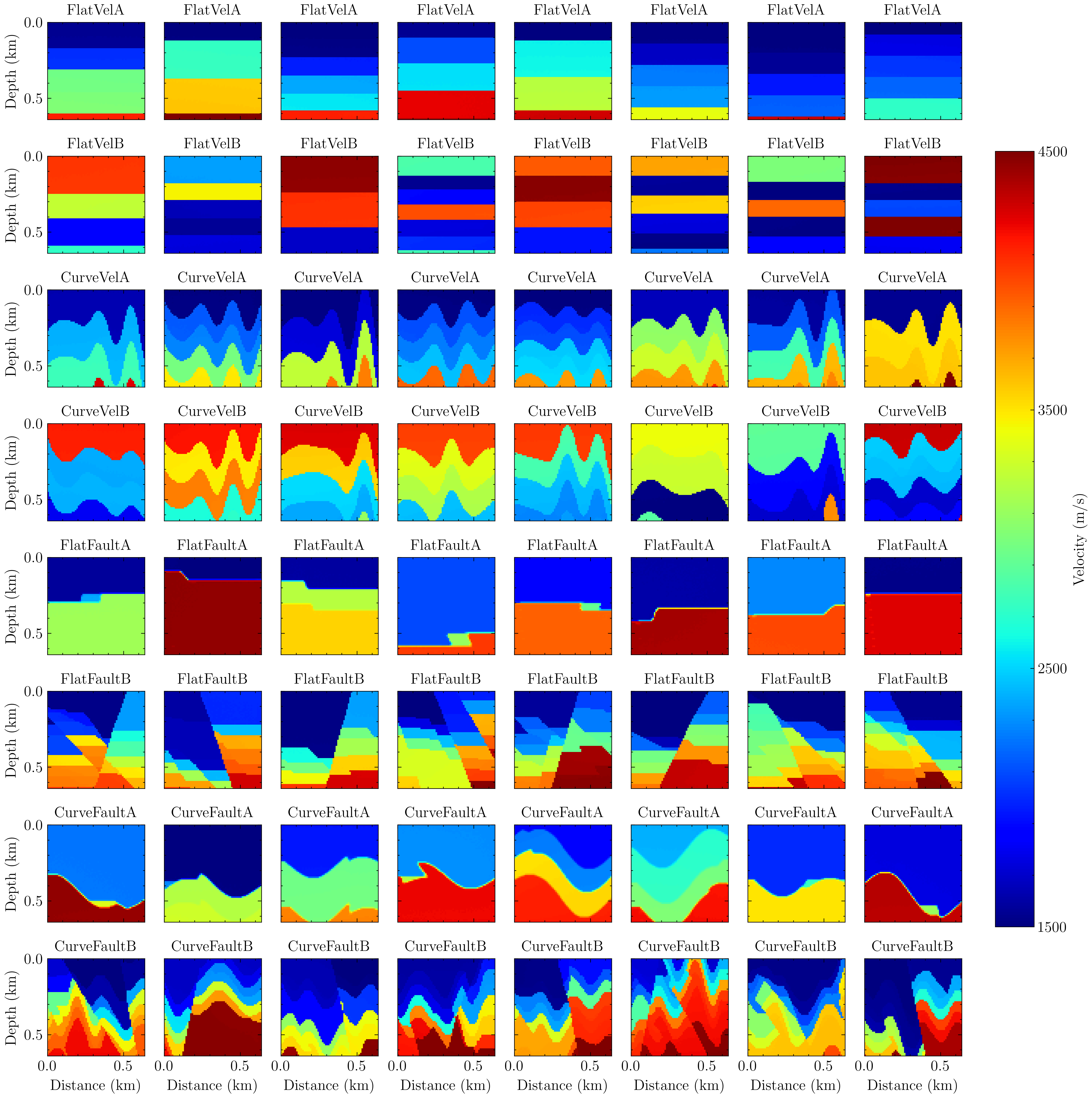}
    \caption{The velocity generation with class conditioning.}
    \label{fig:class_conditional_generation}
\end{figure}

In some scenarios, we would want to control the velocity generation based on prior information. This control can be useful in generating datasets for training as well as in the application of full waveform inversion. In this subsection, we first show the application of class-conditioned velocity synthesis. As mentioned earlier, there are eight classes in the velocity dataset. We incorporate this condition by using a classifier-free guidance diffusion model shown in Figure~\ref{fig:digram2}. For the class embedding, we use a small neural network with one hidden layer of size 256. As suggested in \cite{ho2022classifier}, we jointly train the diffusion model with the condition imposed and without the condition by randomly setting the condition to zero with a probability of 0.5. The sampled results corresponding to their given classes are shown in Figure~\ref{fig:class_conditional_generation}. We observe that the velocity synthesis can be controlled with the class labels, which is favorable to guide the diffusion model in generating the velocity we want. However, this also means the diversity of the velocity synthesis is limited within a given class.

\begin{figure}[h]
    \centering
    \includegraphics[width=1.0\textwidth]{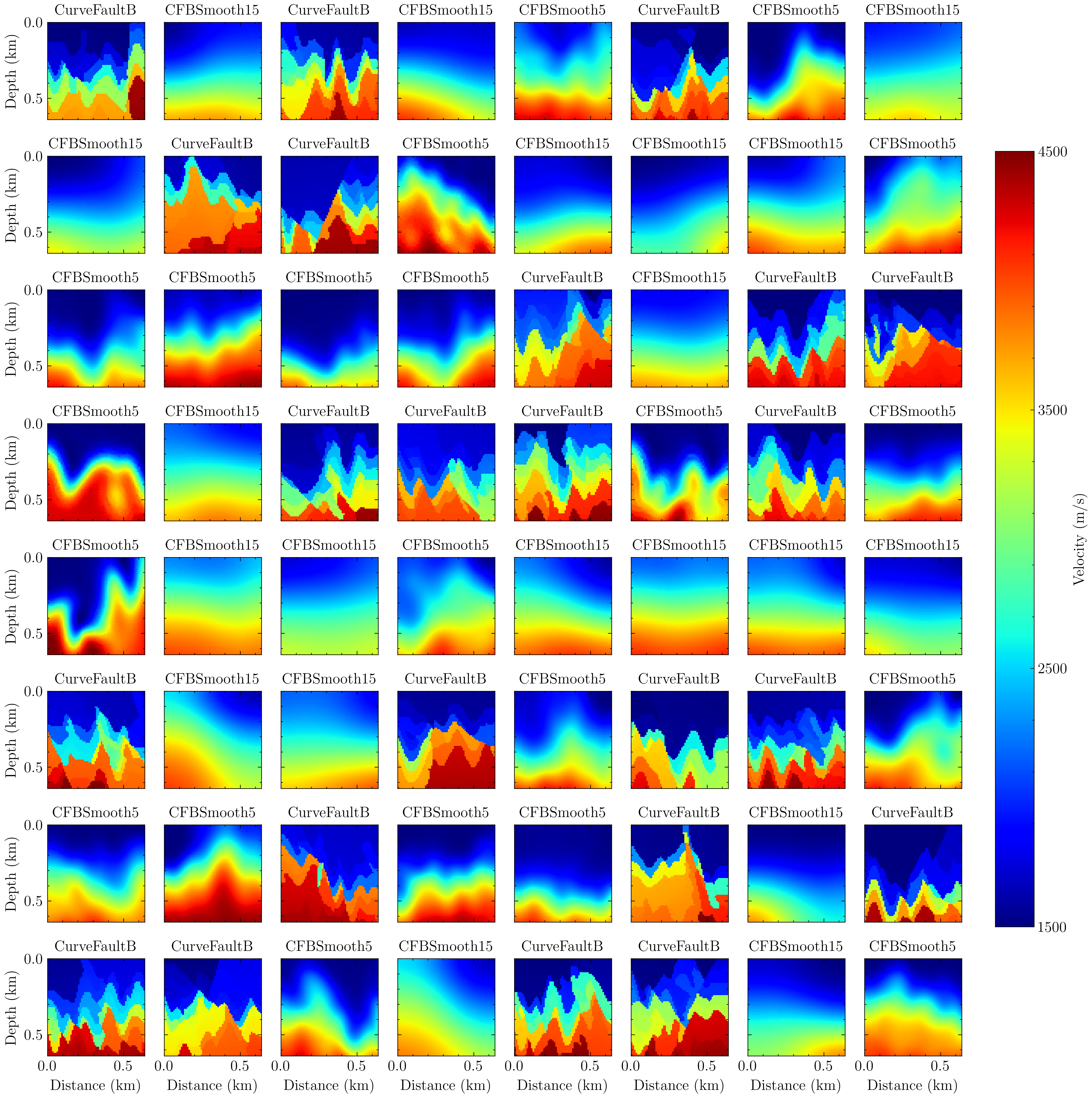}
    \caption{Examples of velocity models corresponding to different smoothness classes.}
    \label{fig:smooth_conditional_generation}
\end{figure}

Another potential class option may involve the smoothness of the velocity model. This can be useful in full waveform inversion since we build the velocity model from smooth to sharp (high-resolution) components. 
Such a conditional generation provides the potential to control the smoothness of the generated velocity. It can help us incorporate the conditional diffusion models into diffusion regularized FWI, in which we can adjust the prior from the diffusion models to align with the model resolution (wavenumber) expected from FWI. Thus, using the smoothness to control the velocity generation would be interesting. Here, to demonstrate the feasibility of the proposed method for such a scenario, we create three classes, from the original ”CurveFaultB” class, based on the level of smothness, including "CurveFaultB", "CurveFaultBSmooth5 (CFBSmooth5)", "CurveFaultBSmooth15 (CFBSmooth5)". "CFBSmooth5", for example, denotes the generated smoothed velocity by using a Gaussian filter applied to the original "CurveFaultB" model with a sigma of 5. The generated results are shown in Figure~\ref{fig:smooth_conditional_generation}, and we can see that the proposed method can handle classifying such cases. Setting a target level of smoothness of the velocity model, the diffusion model generates a velocity with the right amount of smoothness. These generated smooth models correspond to the original distribution, which includes the smoother versions. In future work, we will incorporate such a diffusion model into the workflow of full waveform inversion.

\subsection{Well-conditioned generation}
\label{Well-conditioned generation}
The main purpose of the controllable velocity synthesis is not only for the training dataset generation but also for the velocity generation in an inverse problem \cite[]{wang2023prior}. As commonly used and important prior information that is often available is the well-log information. We usually use the well to generate random velocities to train the neural network, which is a form of incorporating the well as prior information since the distribution is guided by the well. While this type of training may incorporate the prior information from the well itself to guide the training models, it will not constrain the velocity at the well location by its values. Here, we hope to use the well-log information as a constraint, or otherwise control the final generated velocity models to incorporate the well information. That is to say, we hope to generate velocity models that both satisfy the distribution of the velocity models used in the pretraining as well as satisfy the well velocities at the well location.
As mentioned earlier, there are two ways to achieve this goal.

\begin{figure}[h]
    \centering
    \includegraphics[width=1.0\textwidth]{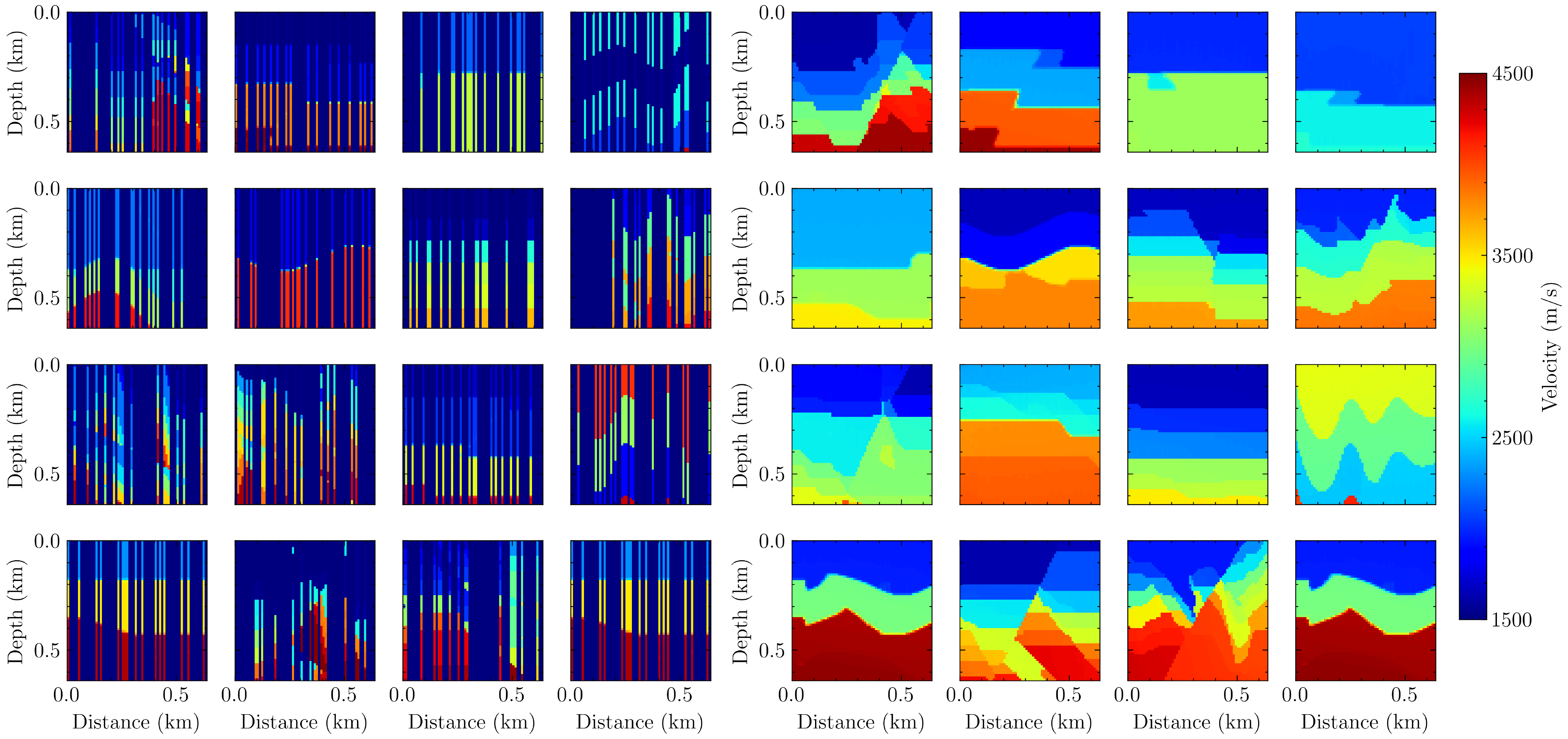}
    \caption{The controllable samples (\textbf{right}) with reconstruction-guided diffusion models, where the condition scale is 1.0 and we use 16 wells (\textbf{left}) as conditions.}
    \label{fig:well-reconstruction-guided-scale1}
\end{figure}

\begin{figure}[h]
    \centering
    \includegraphics[width=1.0\textwidth]{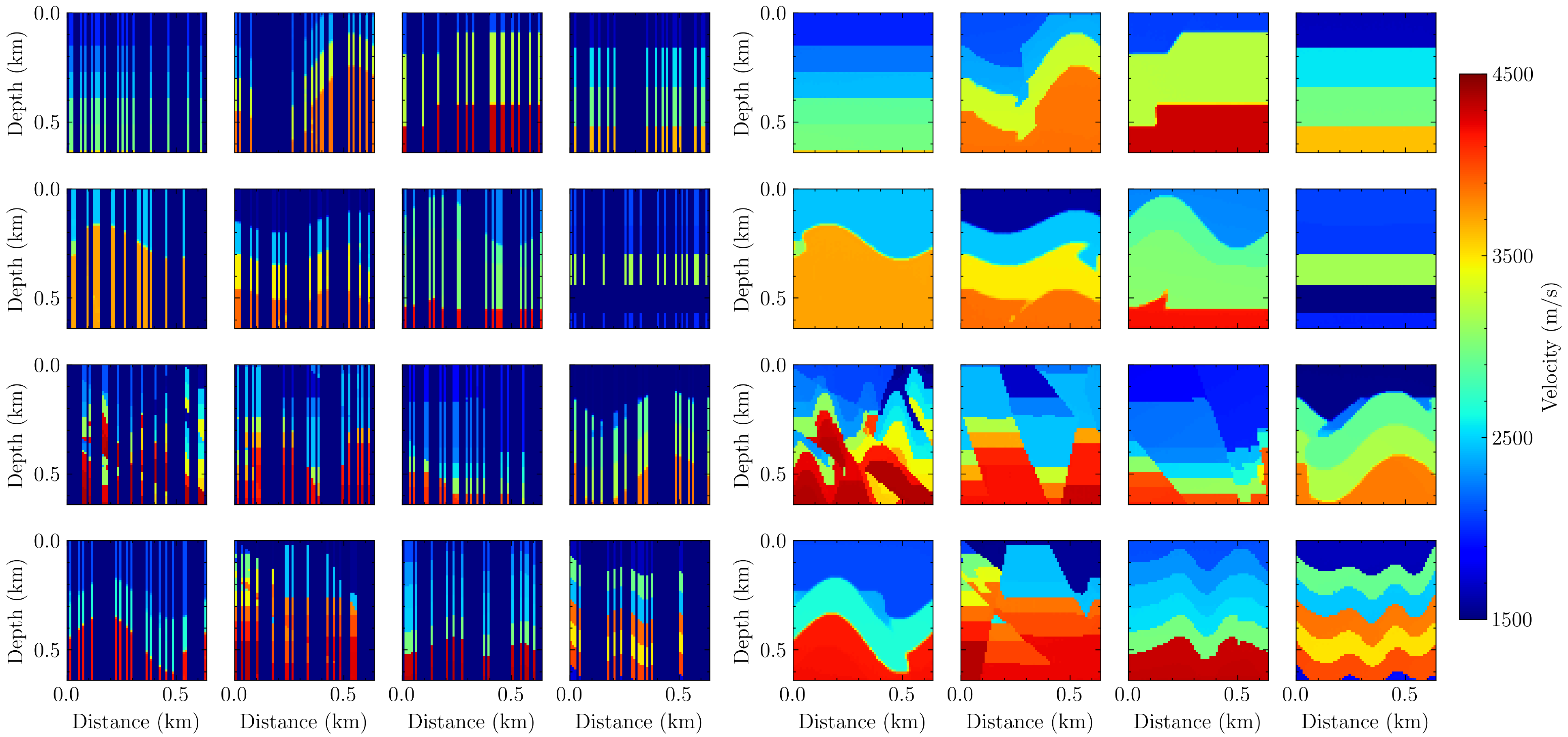}
    \caption{The controllable samples (\textbf{right}) with reconstruction-guided diffusion models, where the condition scale is 8.0, and we use 16 wells (\textbf{left}) as conditions.}
    \label{fig:well-reconstruction-guided}
\end{figure}

\begin{figure}[h]
    \centering
    \includegraphics[width=1.0\textwidth]{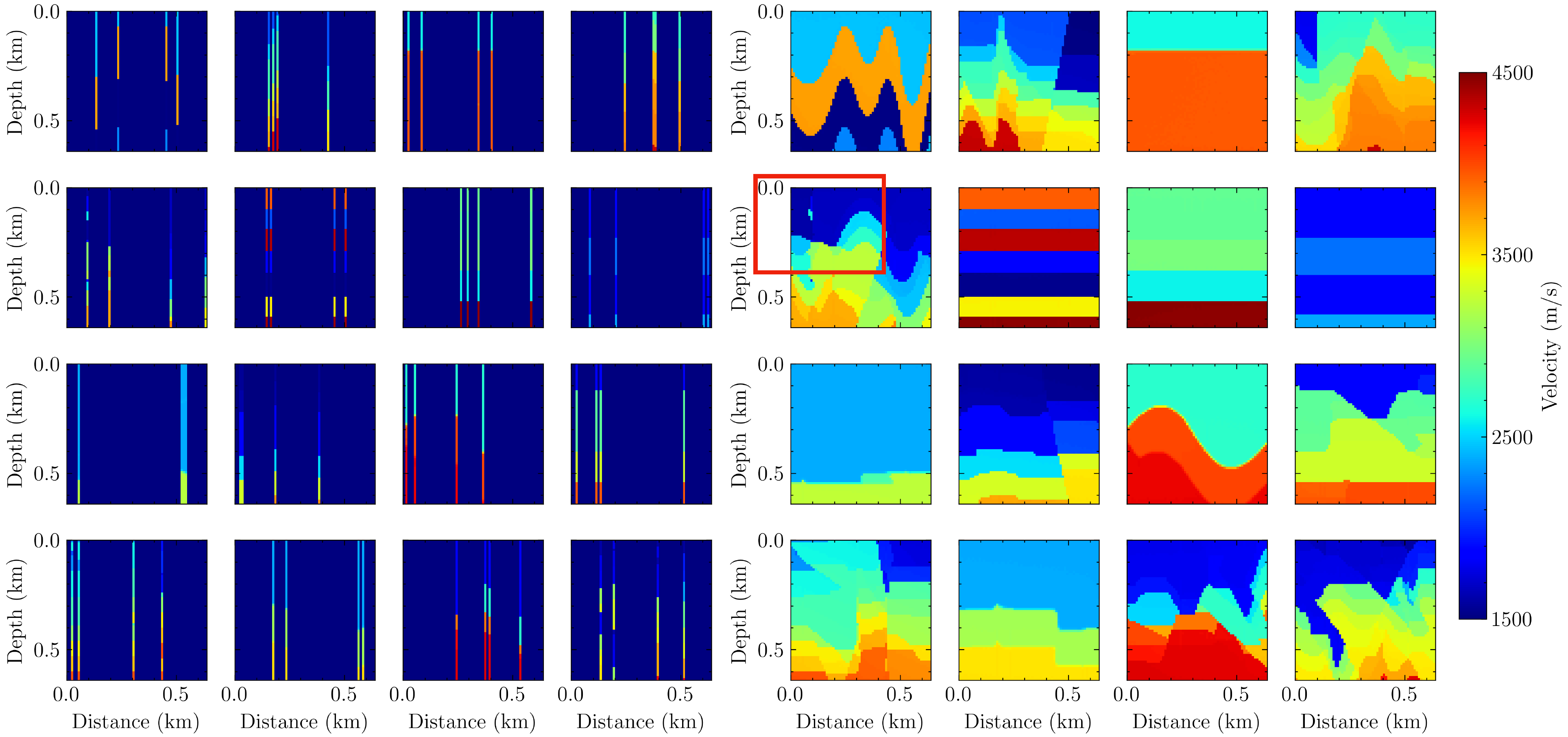}
    \caption{The controllable samples (\textbf{right}) with reconstruction-guided diffusion models, where the condition scale is 32.0, and we use 4 wells (\textbf{left}) as conditions.}
    \label{fig:well-reconstruction-guided-4wells}
\end{figure}

The first one is the reconstruction-guided generation, shown in Figures~\ref{fig:well-reconstruction-guided-scale1} and~\ref{fig:well-reconstruction-guided}. We found that the scale of the condition affects the quality of the results, which is consistent with our previous analysis, where a large condition scale will result in an accurate conditional generation. 
However, when the number of well logs is reduced, we found that the generation is unstable. As shown in Figure~\ref{fig:well-reconstruction-guided-4wells}, although the generated velocities are still consistent with the given wells at the well location, they are poor in structure and horizontal continuity. As denoted by the red box in Figure~\ref{fig:well-reconstruction-guided-4wells}, there are small anomalies. This is caused by the gradient from the reconstruction loss between the well logs and generated velocities, which is insufficient to guide the sampling process to the target that satisfies the well logs as well as the prior distribution. In real-world applications, there are often only a few well-logs in the region, which means this approach may fail to produce an ideal velocity.

\begin{figure}[h]
    \centering
    \includegraphics[width=1.0\textwidth]{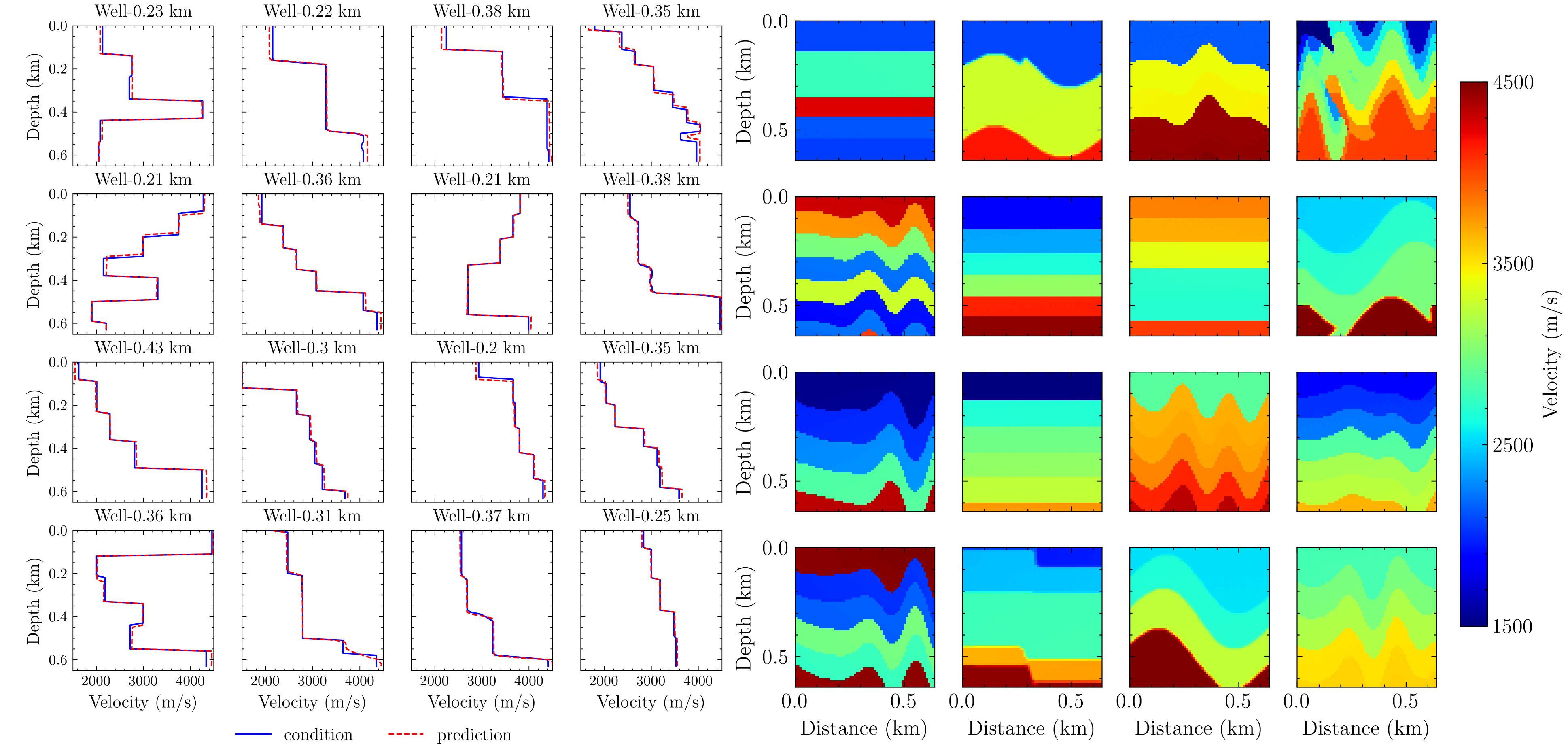}
    \caption{The controllable samples (\textbf{right}) with a classifier-free diffusion model, where we use one well as the condition, and the profile comparison at the well location (\textbf{left}).}
    \label{fig:well-classifier-free-guided}
\end{figure}

\begin{figure}[h]
    \centering
    \includegraphics[width=1.0\textwidth]{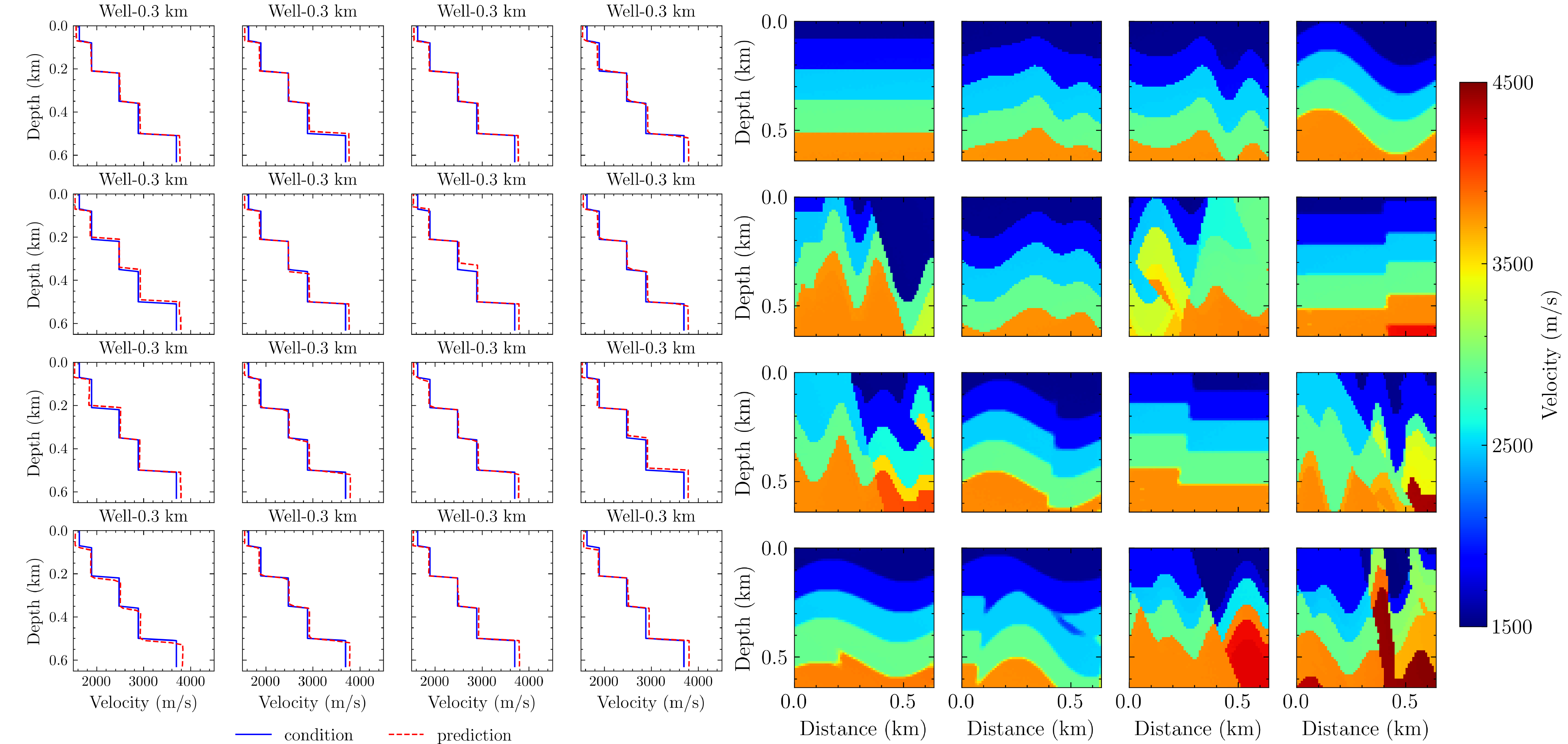}
    \caption{The controllable samples with classifier-free diffusion models, where we use one well located at the horizontal location of 30 as a condition to generate 16 samples (\textbf{right}) to demonstrate the diversity of the generation, and the profile comparison at the well location (\textbf{left}).}
    \label{fig:one-well-classifier-free-guided}
\end{figure}

The second one is the classifier-free generation. As mentioned earlier, this approach implicitly uses the gradient of the neural networks, whose inputs are the well logs, to guide the sampling process. However, it is more robust compared to the reconstruction-guided. In this approach, we can handle the case of a single well log, which is common in realistic problems in 2D surveys. As shown in Figure~\ref{fig:well-classifier-free-guided}, with only one well to guide the sampling process, the generated velocity matches the well log at the well location. Compared to the reconstruction-guided velocity synthesis, this type of method provides a more feasible and more realistic scenario as we usually have one well in the region. To further test the diversity of the proposed method for this scenario, we show 16 samples generated by the same well log in Figure~\ref{fig:one-well-classifier-free-guided}. As expected, the well imposes its information mainly at the location of the well, and away from the well, we are free to generate random models from the distribution. 
Next, to test our method further on other types of conditions, we will show results corresponding to using the structural information to control the velocity synthesis.

\subsection{Seismic-image-conditioned generation}
\begin{figure}[h]
    \centering
    \includegraphics[width=\textwidth]{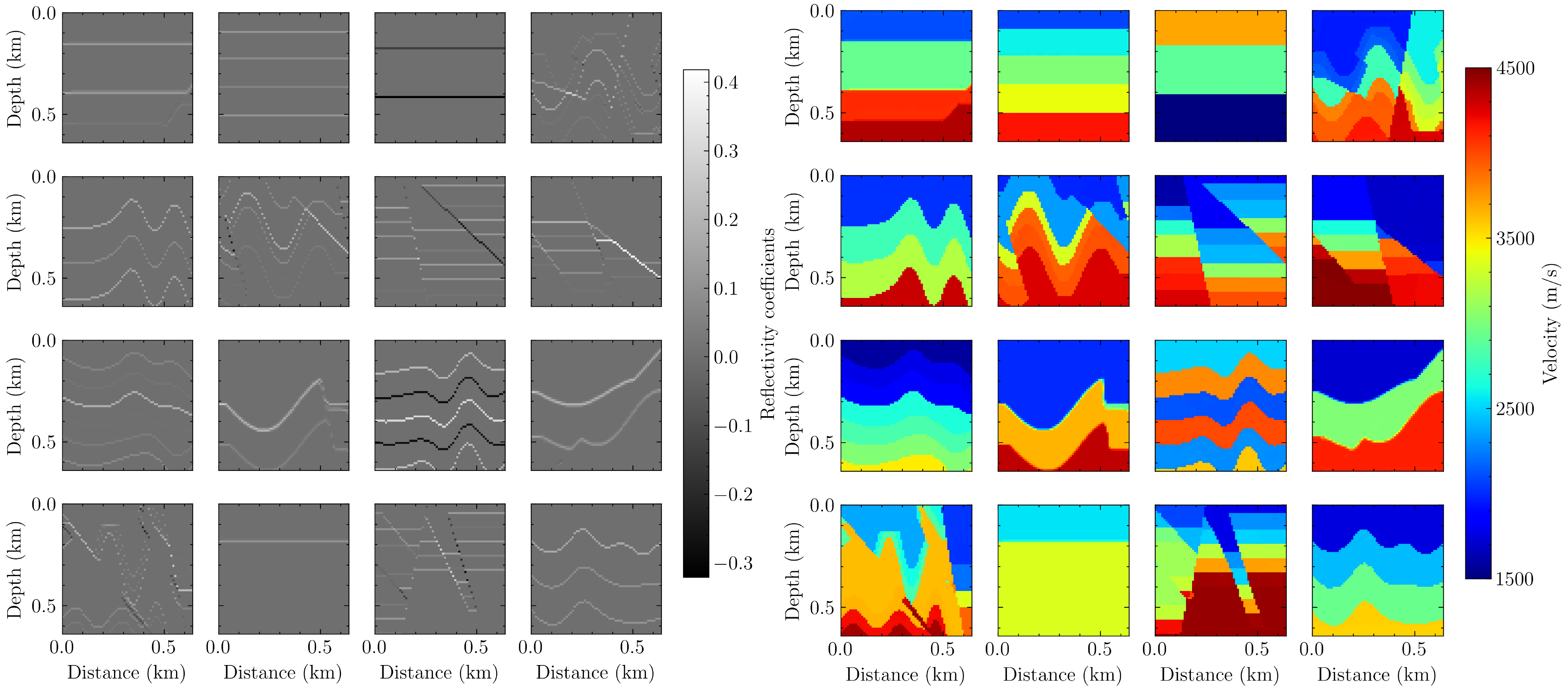}
    \caption{The controllable samples (\textbf{right}) conditioned by the corresponding reflectivity (\textbf{left}) via a cross-attention mechanism.}
    \label{fig:structure-guided-samples}
\end{figure}

In the realm of velocity model generation for data-driven machine learning, as well as its subsequent application in solving inverse problems, one viable approach for exerting control over velocity is through the utilization of images resulting from sketching or migration (exemplified by reflectivity in our paper). Such information allows us to guide the structure of the generated velocity models, aligning them with the desired structural characteristics. It's important to note that, unlike well logs or velocity classes, which are relatively low-dimensional data, images (2D or 3D) contain a wealth of information. Consequently, introducing such desired conditions into the sampling process using a one-hidden-layer MLP (Multi-Layer Perceptron) can be challenging. In response to this challenge, we employ cross-attention mechanisms as an effective strategy to handle the larger dimension of the input conditions given by images. As shown in \cite{rombach2022high}, the cross-attention mechanism used to incorporate such a condition allows for pixel-level control, which can guide the structure of the velocity model to the desired one. Thus, we use cross-attention to add this type of condition. In addition, we use positional information to enhance the accuracy of the control. As shown in Figure~\ref{fig:structure-guided-samples}, the structural information can guide the velocity synthesis, and the reflectivity images are almost consistent with the generated velocities.

\begin{figure}[!htb]
    \centering
    \includegraphics[width=\textwidth]{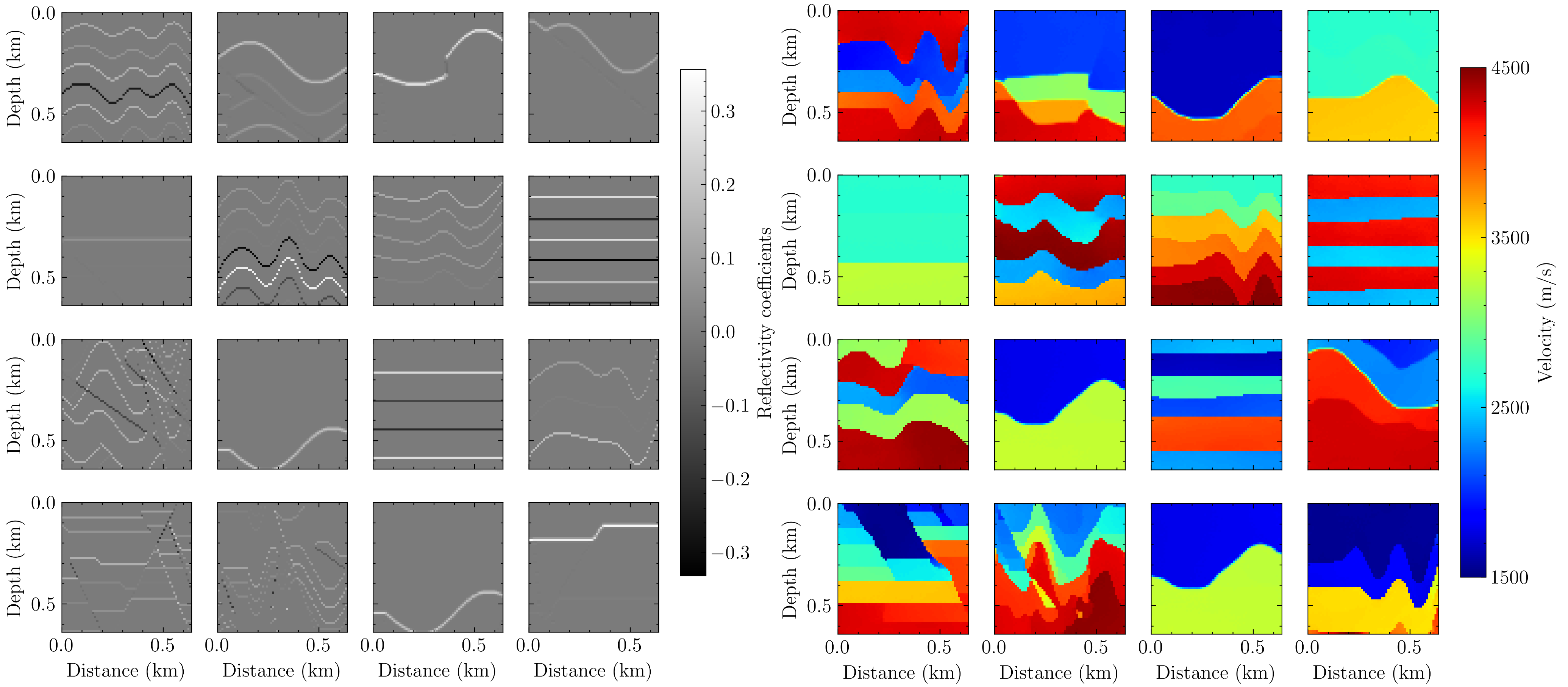}
    \caption{The controllable samples (\textbf{right}) conditioned by the corresponding reflectivity (\textbf{left}) via a cross-attention mechanism without spatial positional encoding.}
    \label{fig:structure-guided-samples-without-pe}
\end{figure}

As mentioned earlier, spatial positional encoding is quite crucial for accurate velocity synthesis control. Here, we show in Figure~\ref{fig:structure-guided-samples-without-pe} cases where the generator fails to adhere to the condition when spatial positional encoding is not used. We observed that although the generated velocities have similar structural information to the given reflectivity, they are not the same. In some cases, they are out of symmetry with the reflectivity image. This is because the cross-attention can only capture the structural information without the proper positioning or direction. The spatial positional encoding mitigates this problem by injecting positional information on the reflectivity into the diffusion model, yielding more accurate controllable velocity synthesis.

\subsection{Integrated all-conditions for generation}
\begin{figure}[h]
    \centering
    \includegraphics[width=\textwidth]{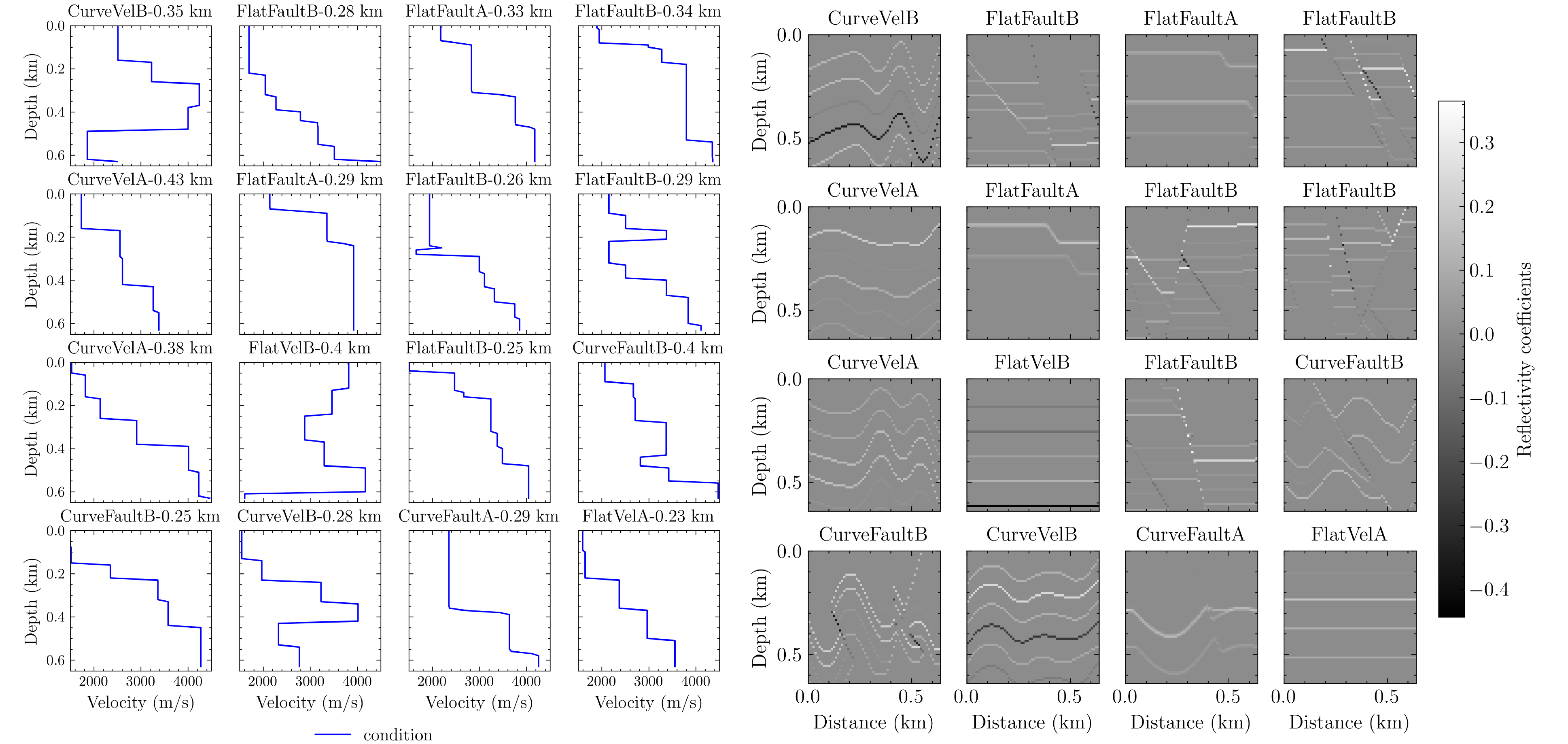}
    \caption{The well logs (\textbf{left}) and reflectivity (\textbf{right}) used for controlling the velocity generation.}
    \label{fig:well_reflectivity}
\end{figure}

\begin{figure}[!htb]
    \centering
    \includegraphics[width=\textwidth]{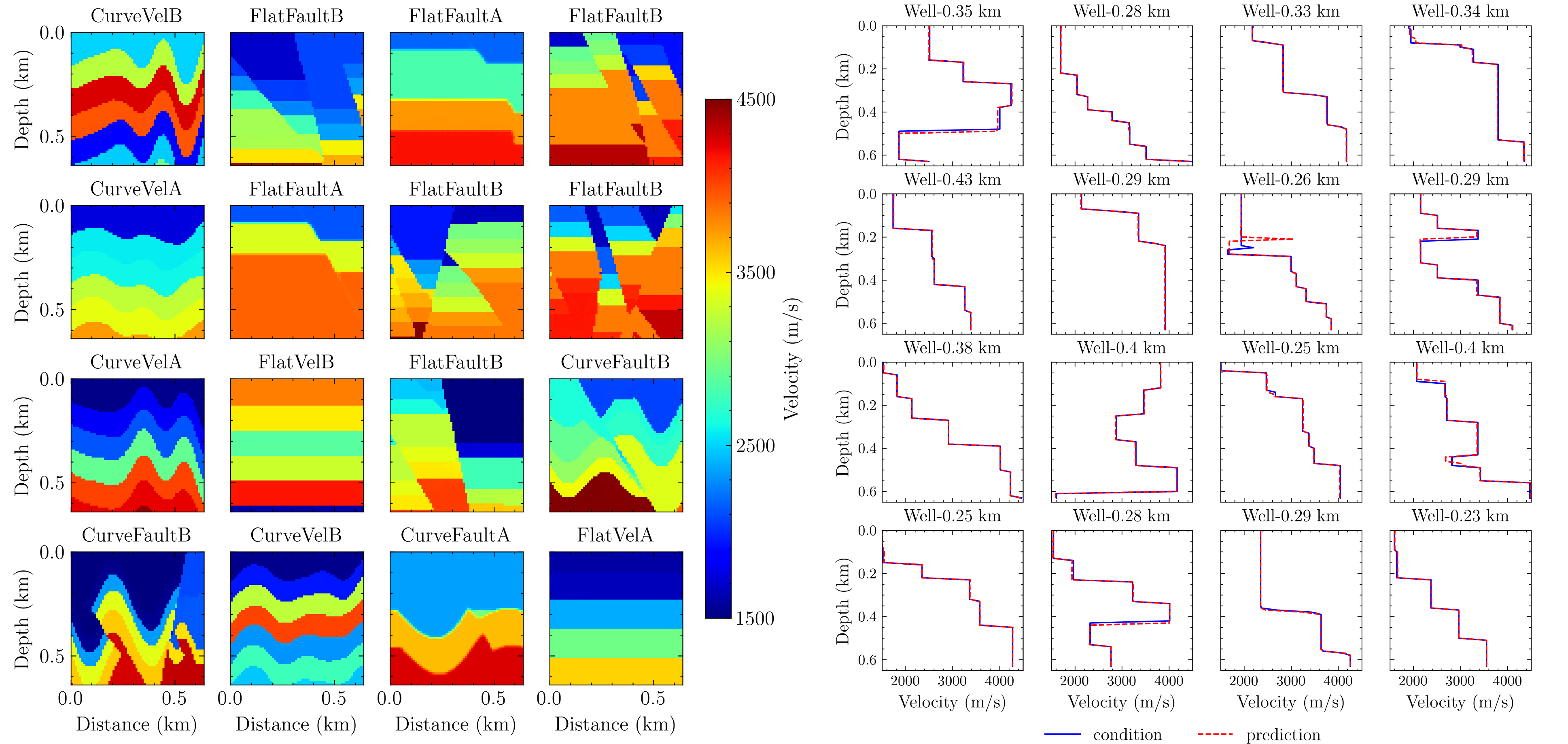}
    \caption{The controllable samples (\textbf{left}) conditioned using class labels, well logs, and reflectivity images, and the profile comparisons (\textbf{right}) between the well log and generated the velocities at the well location.}
    \label{fig:integrated_conditions_samples}
\end{figure}

The above examples demonstrated the conditional abilities to guide the generation of velocities. 
In this subsection, we test our approach with the integration of all relevant conditions to exercise precise control over the velocity generation. In Figure~\ref{fig:well_reflectivity}, we can observe the inclusion of well logs and reflectivity images alongside the accompanying class labels, serving as input conditions. The resulting velocity samples, as depicted in Figure~\ref{fig:integrated_conditions_samples}, demonstrate our ability to generate velocities that closely align with the desired criteria. Notably, we have also extracted velocity profiles at the precise well-log locations, and our findings reveal that, in the majority of cases, these profiles maintain a high degree of fidelity and consistency with the intended parameters. 

\section{Discussion}
Unlike GANs, diffusion models require hundreds of iterative steps of reverse diffusion to generate synthetic velocities. Thus, it is computationally slower at the inference stage, which is a known drawback of Diffusion models.  This drawback is the trade-off for the richness and high quality of the generated images produced by the diffusion model. Nevertheless, the slower generation is being addressed via the denoising diffusion implicit models \cite[]{song2020denoising} or operator learning for fast sampling \cite[]{zheng2023fast}. Thus, considering its stable training and high-quality generation, compared to GANs, many have tolerated the slow generation in favor of quality \cite[]{ramesh2022hierarchical}. In terms of application to FWI, \cite{wang2023prior} have shown that incorporating the diffusion model into the FWI will only add negligible computational cost.

\begin{figure}[h]
    \centering
    \includegraphics[width=1.0\textwidth]{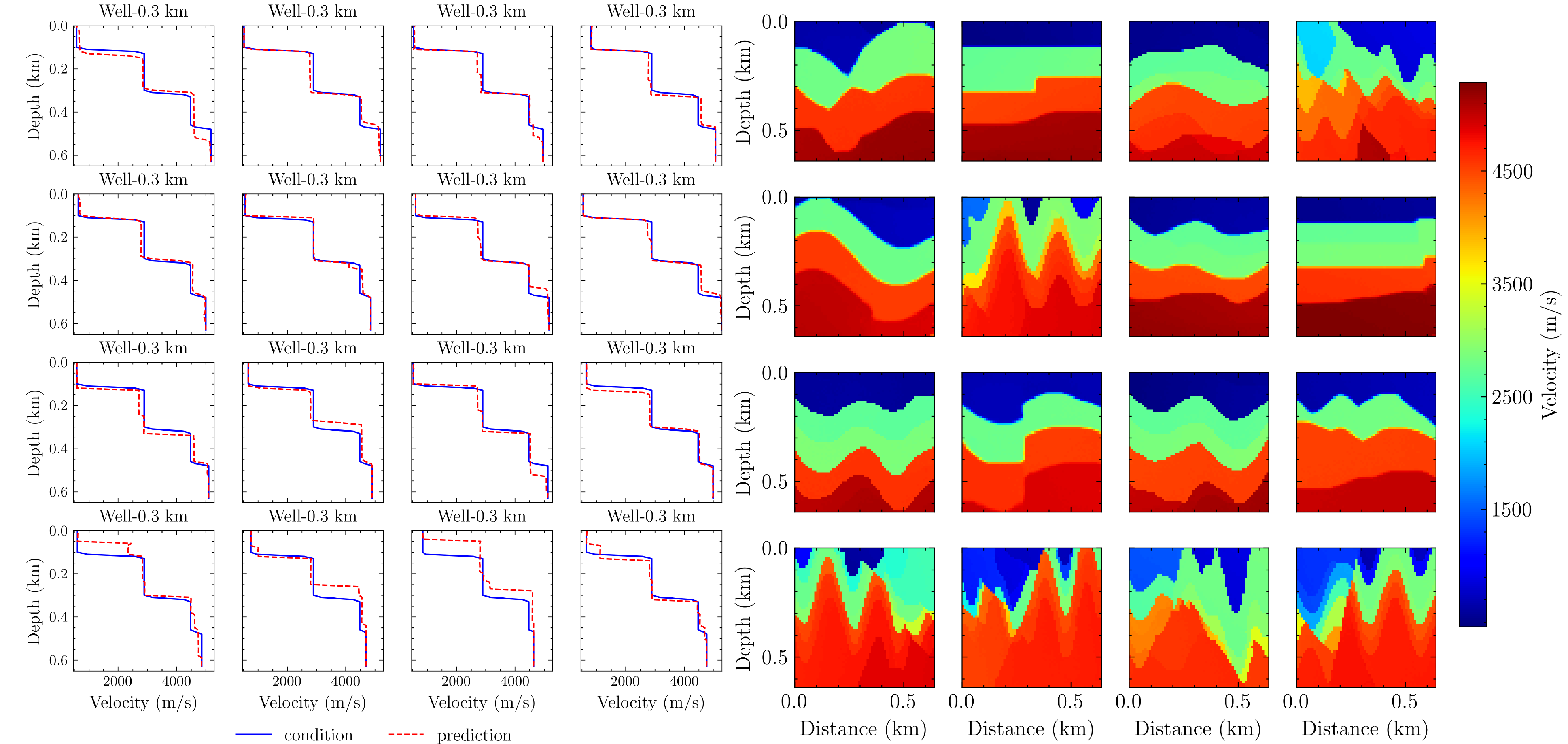}
    \caption{The controllable samples (\textbf{right}) with classifier-free diffusion model test using an out-of-distribution sample, with velocities out of the range of the training set. This well (\textbf{left}) is located at the horizontal location of 30 as a condition to generate 16 samples to demonstrate the diversity of the generation.}
    \label{fig:one-well-classifier-free-guided-ood}
\end{figure}

\begin{figure}[!htb]
    \centering
    \includegraphics[width=1.0\textwidth]{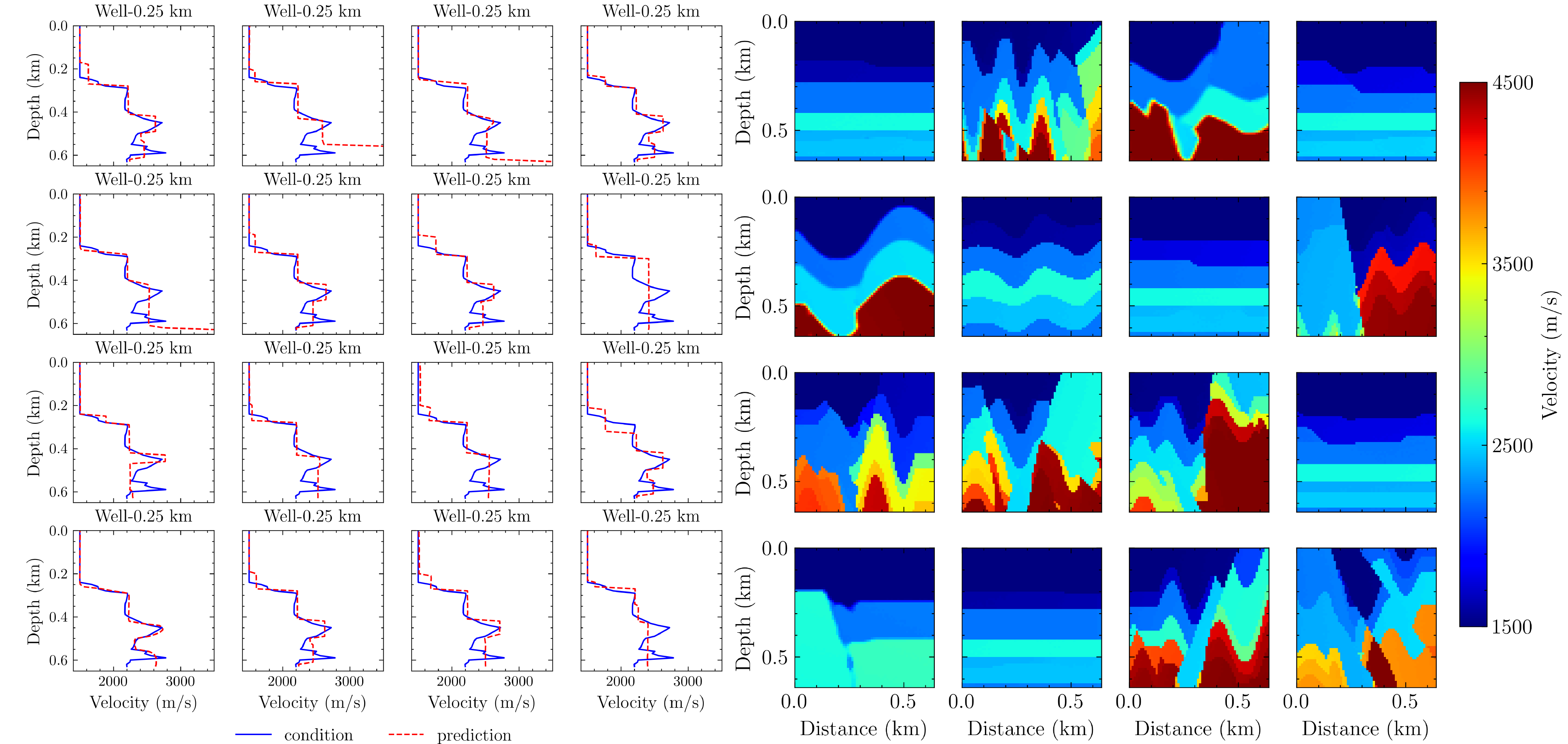}
    \caption{The controllable samples (\textbf{right}) with a classifier-free diffusion model using another out-of-distribution condition from field data. The location of the well (\textbf{left}) is at 25, and it corresponds to a 2D field marine dataset from the North-Western Australia Continental shelf acquired by CGG to generate 16 samples to demonstrate the diversity.}
    \label{fig:one-well-classifier-free-guided-ood-cgg}
\end{figure}

Another important question that may arise is how well these generatations work with out-of-distribution conditions. To test this, we focus our attention on the well-condition implementation shared in Section~\ref{Well-conditioned generation}.  
In realistic scenarios, we may face the issue of out-of-distribution controllable velocity synthesis as the well velocities may exceed the minimum or maximum velocity of the training dataset.
Here, we found that the proposed method can still handle such out-of-distribution cases, as shown in Figure~\ref{fig:one-well-classifier-free-guided-ood}.
Although the generated velocities do not perfectly match the well due to its out-of-distribution nature, where the maximum velocity is 5250 $m/s$, and the minimum velocity is 460 $m/s$, it still provides a good approximation of the well velocity. 
To further demonstrate the performance of our conditional velocity synthesis, we use the Maximum Mean Discrepancy (MMD) \citep{smola2006maximum} to quantify the difference between the original distributions and the distribution of the generated velocities. For the MMD calculation, we randomly sampled 1000 velocities from the original training dataset while generating 64 samples from out-of-distribution well logs. The MMD for this out-of-distribution test is 2.6454. With the same setting, the in-distribution test often results in a value less than 1.0. Because a large value of MMD denotes a larger distance between two distributions, the generated velocities are far away from the training distribution.
Another out-of-distribution test is a continuously changing well velocity extracted from the field, and such rapid velocity changes are not represented in our trained models (see Figure~\ref{fig:unconditional_generation}). The well log is located at the North-Western Australia Continental shelf and is acquired by CGG, and the resulting generated velocity models are shown in Figure~\ref{fig:one-well-classifier-free-guided-ood-cgg}. We can see that the general velocity profiles from the generated results match the main trend of the well velocity, and the corresponding MMD is 1.4221.
This demonstrates the potential application for realistic controllable velocity generation. Of course, if the diffusion model were trained on velocity models resembling this well, then the match would be stronger.

\begin{figure}[h]
    \centering
    \includegraphics[width=1.0\textwidth]{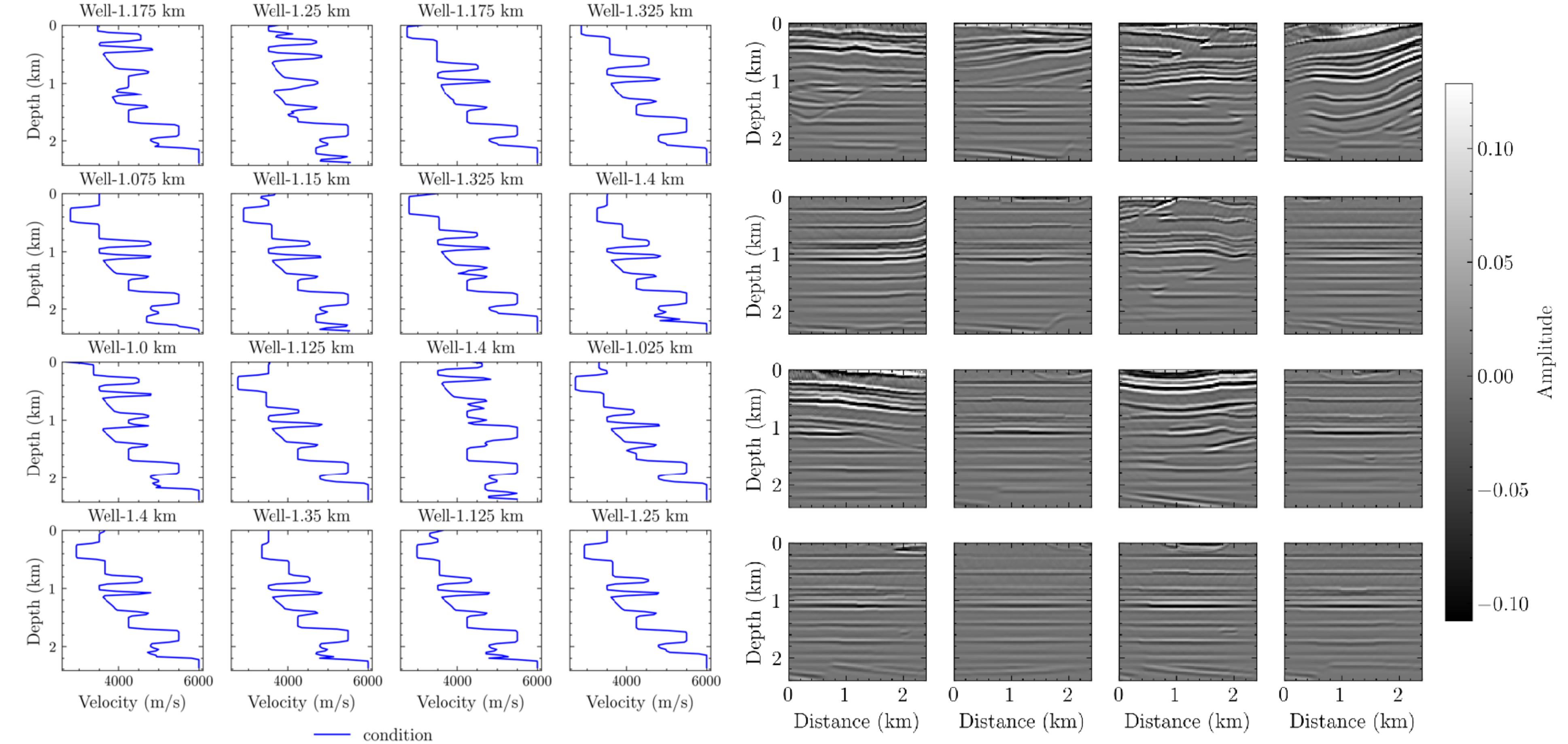}
    \caption{The well logs (\textbf{left}) and RTM results (\textbf{right}) used for controlling the velocity generation.}
    \label{fig:overthrust-condition}
\end{figure}

\begin{figure}[!htb]
    \centering
    \includegraphics[width=1.0\textwidth]{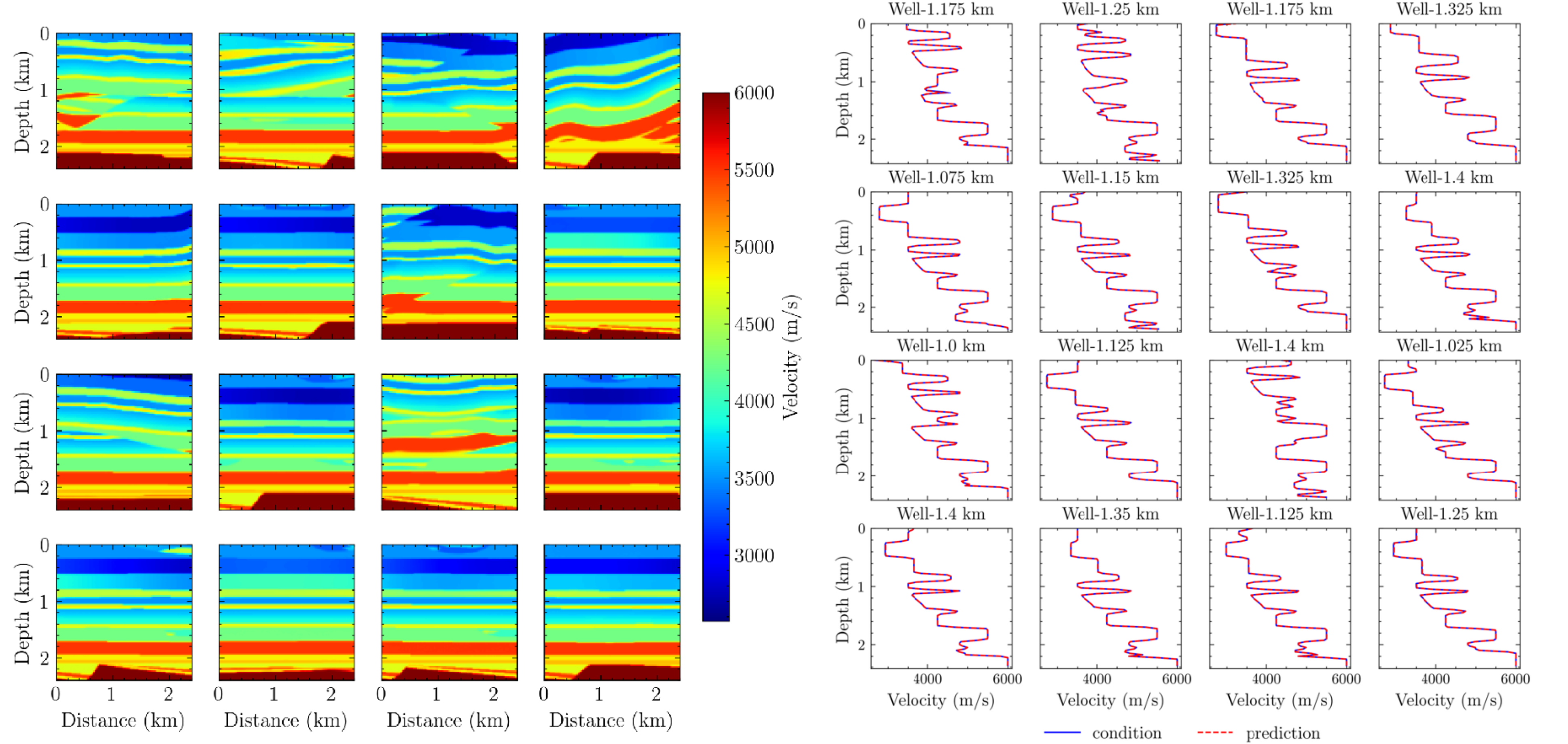}
    \caption{The generated samples (\textbf{left}) conditioned using well logs and RTM results, and the profile comparisons (\textbf{right}) between the well log and the generated velocities at the well location.}
    \label{fig:overthrust-generation-comparison}
\end{figure}

With respect to image-conditioned generation, as mentioned earlier, we can alternatively use the imaging results from applying reverse time migration (RTM) using a smooth version of the velocity models (possibly during training) or those obtained from migration velocity analysis.
Thus, we extracted the 2000 velocity patches from the 3D Overthrust model, which has a reasonably complex geology and whose velocity varies spatially at the pixel level, to test the proposed method
For such a limited number of velocity samples (2000) and limited diversity (extracted from one model), the resulting trained diffusion model is expected to suffer from poor generalization.
The purpose here is to demonstrate the proposed method in controllable velocity generation using well logs and RTM images. 
The size of each velocity patch is 96$\times$96 with an interval of 25 $m$ in both vertical and horizontal directions. Though these are patches extracted from a 3D velocity model, we assume they represent individual velocity models for the same 2D area, as we aim to train the diffusion model to store their distribution.
We smooth each velocity model with a Gaussian filter and then apply RTM to obtain the corresponding seismic imaging results, utilizing a 24Hz ricker wavelet and 10 sources, followed by a Laplacian filter.
We use the same training configuration as that for the OpenFWI test but change the input resolution to 96$\times$96 and include the RTM images as well as random well logs as conditions.
Given unseen well logs and RTM images from the test set (Figure \ref{fig:overthrust-condition}), our proposed method can synthesize velocity models that align with the RTM images and correlate well with the well log, as demonstrated in Figure \ref{fig:overthrust-generation-comparison}. 
This simple example demonstrates the effectiveness of the proposed method in such a scenario.

Overall, the proposed method can handle multiple conditions in the velocity synthesis.
However, in realistic applications, sometimes we may have conditions beyond the types studied in this paper. The proposed method, especially the use of cross-attention, can handle multiple multi-modal conditions inherently. We note that if the samples with conditions used for training are not sufficient, the proper way to incorporate the other conditions with limited samples (e.g., expert manual interpretation of the subsurface structures might be limited) is to use ControlNet \cite[]{zhang2023adding}. With this method, we can lock the pre-trained diffusion model and reuse the encoding layers as a backbone to learn diverse conditional controls, whose samples are not sufficient, to add more conditions.

\section{Conclusions}
\label{conclusion}
We presented a controllable velocity synthesis framework using diffusion models, yielding a high-diversity and high-quality velocity synthesis. 
We show different ways to control the velocity synthesis, e.g., reconstruction-guided sampling and classifier-free guidance diffusion models.
We found that the classifier free guidance diffusion model can stably generate the velocity given the target conditions, including single or integrated multiple conditions.
Furthermore, the cross-attention mechanism can handle image-like conditions (such as subsurface structure) and can produce highly accurate velocity generation control.
The proposed method shows great potential to support multi-model prior information incorporation for diffusion-regularized FWI and provide massive and diverse training datasets to augment training for neural network-based inversion algorithms.

\section{Acknowledgments}
We thank KAUST and the DeepWave Consortium sponsors for their support. This work utilized the resources of the Supercomputing Laboratory at KAUST, and we are grateful for that.

\bibliographystyle{plainnat}
\bibliography{agusample}
\end{document}